\RequirePackage{amsmath} 
\documentclass{llncs}

\usepackage{stmaryrd}
\usepackage{microtype} 
\usepackage{amssymb} 
\usepackage{graphicx}
\usepackage[pdftex,dvipsnames]{xcolor}  
\usepackage{algorithm}
\usepackage[noend]{algpseudocode}
\usepackage[T1]{fontenc} 
\usepackage{multirow}
\usepackage{fp} 
\usepackage{hyperref} 
\usepackage{marvosym} 

\newcommand{\FS}{\textsf{FS}}
\newcommand{\MS}{\textsf{MS}}

\newcommand{\defs}[1]{\ensuremath{\mathsf{def}(#1)}}
\newcommand{\cls}[1]{\ensuremath{\mathsf{cls}(#1)}}

\newcommand{\exps}[1]{\ensuremath{\mathsf{exp}(#1)}}
\newcommand{\imps}[1]{\ensuremath{\mathsf{imp}(#1)}}

\newcommand{\mods}[1]{\ensuremath{\mathsf{mod}(#1)}}

\newcommand{\esc}[2]{\ensuremath{\mathsf{esc}_{#1}(#2)}}
\newcommand{\vis}[2]{\ensuremath{\mathsf{vis}_{#1}(#2)}}
\newcommand{\usr}[1]{\ensuremath{\mathsf{usr}(#1)}}

\newcommand{\vars}[1]{\ensuremath{\mathsf{Vars}_{#1}}}

\newcommand{\gd}[0]{\mid}
\newcommand{\state}[2]{\ensuremath{\langle #1 \gd{} #2 \rangle}}

\newcommand{\emptyGoal}{\ensuremath{\square}}

\newcommand{\reduction}[2]{\ensuremath{#1 \leadsto #2}}
\newcommand{\reductionRtc}[2]{\ensuremath{#1 \leadsto_{\mathsf{rtc}} #2}}

\newcommand{\errorErase}[1]{\ensuremath{#1^{\circ}}}

\newcommand{\A}{\ensuremath{{\cal A}}}
\newcommand{\AC}{\ensuremath{{\cal A}_{C}}}

\newcommand{\callsAsr}[2]{\ensuremath{\textsf{calls}(#1, #2)}}
\newcommand{\successAsr}[3]{\ensuremath{\textsf{success}(#1, #2, #3)}}

\newcommand{\asrId}[1]{\ensuremath{c_{#1}}}

\newcommand{\ret}[1]{\ensuremath{\textsf{ret}(#1)}}
\newcommand{\retchk}[2]{\ensuremath{\textsf{ret}(#1,#2)}}

\newcommand{\err}[1]{\ensuremath{\textsf{err}(#1)}}
\newcommand{\errchk}[1]{\ensuremath{\textsf{err}(\asrId{#1})}}

\newcommand{\labCallsAsr}[3]{\ensuremath{\textsf{\asrId{#1}.calls}(#2, #3)}}
\newcommand{\labSuccessAsr}[4]{\ensuremath{\textsf{\asrId{#1}.success}(#2,#3,#4)}}

\newcommand{\answers}[1]{\ensuremath{\textsf{answers}(#1)}}

\newcommand{\Q}{\ensuremath{{\cal Q}}}

\newcommand{\derivations}[1]{\ensuremath{\textsf{derivs}(#1)}}
\newcommand{\derivationsRtc}[1]{\ensuremath{\textsf{rtc-derivs}(#1)}}

\newcommand{\kbd}[1]{\mbox{\tt #1}}

\newcounter{myexamplecounter}
\newenvironment{myexample}{%
  \refstepcounter{myexamplecounter}%
  \smallskip
  \noindent
  \emph{Example \arabic{myexamplecounter}. \ }%
}{%
}

\newcounter{parentnumber}


\newenvironment{shortenumerate}
{ \setlength{\itemsep}{0pt}
  \setlength{\parskip}{1pt}
  \setlength{\parsep}{0pt}
  \setlength{\topsep}{0pt}
  \setlength{\partopsep}{0pt}
  \begin{enumerate} }
{ \end{enumerate}   }

\newcommand{\finalcompression}[1]{\vspace*{#1}}

\begin{document}

\title{Exploiting Term Hiding to Reduce Run-time~Checking~Overhead
  \thanks{%
     This work has been submitted and accepted to PADL'18. %
     An extended abstract of this work is published
     as~\cite{termhide-iclp2017-tc-short}}}

\author{Nataliia Stulova\Letter\inst{1,2}
        \and
        Jos\'{e} F. Morales\inst{1}
        \and
        Manuel V. Hermenegildo\inst{1,2}
}

\institute{%
IMDEA Software Institute, Madrid, Spain \\
\and %
ETSI Inform\'{a}ticos, Universidad Polit\'{e}cnica de Madrid (UPM), Madrid, Spain \\
\email{\texttt{\{nataliia.stulova,josef.morales,manuel.hermenegildo\}@imdea.org}}
}


\maketitle

\finalcompression{-7mm}
\begin{abstract}
One of the most attractive features of untyped languages
is the flexibility in term creation and manipulation.
However, with such power comes the responsibility of ensuring the
correctness of these operations.
A
solution is adding run-time checks to the program via assertions, but this
can introduce overheads that are in many cases impractical.
While static analysis can greatly reduce such overheads, the gains
depend strongly on the quality of the information inferred.
Reusable libraries, i.e., library modules that are pre-compiled
independently of the client, pose special challenges in this context.
We propose a technique which takes advantage of module systems which can
hide a selected set of functor symbols to significantly enrich the shape
information that can be inferred for reusable libraries,
as well as an improved run-time checking approach that leverages the
proposed mechanisms to achieve large reductions in overhead, closer to
those of static languages, even in the reusable-library context.
While the approach is general and system-independent, we present it
for concreteness in the context of the Ciao assertion language and
combined static/dynamic checking framework.
Our method maintains the full expressiveness of the assertion language
in this context.
In contrast to other approaches it does not introduce the need to switch
the language to a (static) type system, which is known to change the
semantics in languages like Prolog.
We also study the approach experimentally
and evaluate the overhead reduction achieved in the run-time checks.
\end{abstract}

\finalcompression{-4mm}


\finalcompression{-3mm}
\section{Introduction}

Modular programming has become widely adopted
due to the benefits it provides
in code reuse and
structuring data flow between program components.
A tightly related concept is the principle of \emph{information hiding}
that allows concealing the concrete implementation details behind a
well-defined interface and thus allows for cleaner abstractions.
Different programming languages implement these concepts in different
ways, some examples being the encapsulation mechanism of classes in
object-oriented programming and opaque data types.
In the (constraint) logic programming context, most mature language
implementations incorporate module systems,
some of which allow programmers to restrict the visibility of some
functor symbols
to the module where they are defined, thus both hiding the concrete
implementation details of terms from other modules and providing
guarantees that only the predicates of that particular module can use
those functor symbols as term constructors or matchers.

One of the most attractive features of untyped languages for
programmers is the flexibility they offer in term creation and
manipulation.  However, with such power comes the responsibility of
ensuring correctness in the manipulation of data, and this is
specially relevant when data can come from unknown clients.
A popular solution for ensuring safety is to enhance the language with
optional assertions that allow specifying correctness conditions both
at the module boundaries and internally to modules.  These assertions
can be checked dynamically by adding run-time checks to the program,
but this can
introduce overheads that are in many cases
impractical.
Such overheads can be
greatly reduced with static analysis, but the gains then depend
strongly on the quality of the analysis information inferred.
Unfortunately, there are some common scenarios where shape/type analyses
are necessarily imprecise.
A motivational example is the case of reusable libraries, i.e., the case
of analyzing, verifying, and compiling a library for general use,
without access to the client code or analysis information on it.
This includes for example the important case of servers accessed via
remote procedure calls.
Static analysis faces
challenges in this context, since the unknown clients can fake data
that is really intended to be internal to the library.
Ensuring safety then requires sanitizing input data with
potentially expensive run-time checks.

In order to alleviate this problem, we present techniques that, by
exploiting term hiding and the strict visibility rules of the
module system, can greatly improve the quality of the shape
information inferred by static analysis
and reduce the run-time overhead for the calls across module
boundaries by several orders of magnitude.
These techniques can result in improvements in the number and size of
checks that allow bringing guarantees and overheads to levels
close
to those of statically-typed approaches, but without imposing on
programs the restriction of being well-typed.
For concreteness, we use in this work the relevant parts of the Ciao
system~\cite{hermenegildo11:ciao-design-tplp-shorter},
which pioneered
the assertion-based, combined static+dynamic checking approach:
the module system,
the assertion language --which allows providing optional program
specifications with various kinds of information, such as modes,
shapes/types, non-determinism, etc.--,
and the overall framework.
However, our results are general and we believe they can be applied to
many dynamic languages.  In particular, we present a semantics for
modular logic programs where the mapping of module symbols is abstract
and implementation-agnostic, i.e., independent of the visibility rules
of particular module systems.


\section{Preliminaries}

We first recall
some basic notation and the standard program semantics, following the
formalization of~\cite{optchk-ppdp2016-shorter}.
An \emph{atom} $A$ is a syntactic construction of the form
$f(t_1,\ldots,t_n)$ where $f$ is a symbol of arity $n$ and the $t_i$
are \emph{terms}.
Terms are inductively defined as variable symbols or
constructions of the form $f(t_1,\ldots,t_n)$ where $f$ is a symbol of
arity $n~(n \geq 0)$ and the $t_i$ are terms.
Note that we do not (yet) distinguish between predicate symbols and
functors (uninterpreted function symbols), denoting the global set of
\emph{symbols} as $\FS$.
A \emph{constraint} is a conjunction of expressions built
from predefined predicates (such as term equations or inequalities
over the reals) whose arguments are constructed using predefined
functions (such as real addition).
A \emph{literal} is either an atom or a constraint.
\emph{Constants} are introduced as 0-ary symbols.
A \emph{goal} is a conjunction of literals.
A \emph{clause} is defined as $H \leftarrow B$, where $H$ is an atom
(the head)
and $B$ is a goal (the body).
A \emph{definite program} is a finite set of clauses.
The \emph{definition} of an atom $A$ in a program, $\cls{A}$, is the
set of program clauses
whose head has
the same predicate symbol and arity as $A$, renamed-apart.
We assume that all clause heads are \emph{normalized}, i.e., $H$ is of
the form $f(X_1,\ldots,X_n)$ where the $X_1,\ldots,X_n$ are distinct
free variables.

We
recall
the classic operational semantics of
(non-modular) definite programs,
given in terms of program \emph{derivations}, which are sequences of
\emph{reductions} between \emph{states}.
We use $::$ to denote concatenation of sequences.
A \emph{state} $\state{G}{\theta}$ consists of a goal sequence $G$ and a
constraint store (or \emph{store} for short) $\theta$.
A \emph{query} is a pair $(L,\theta)$, where $L$ is a literal and
$\theta$ a store, for which the (constraint) logic programming system
starts a computation from state $\state{L}{\theta}$.
The set of all derivations from the query $Q$ is denoted
$\derivations{Q}$.
A finite derivation from a query $(L,\theta)$ is
\emph{finished} if the last state in the derivation cannot be
reduced,
and it is
\emph{successful} if the last state is of the form
$\state{\emptyGoal}{\theta'}$, where $\emptyGoal$ denotes the empty
goal sequence.
In that case, the constraint $\bar{\exists}_{L}\theta'$
(denoting the projection of $\theta$ onto the variables of $L$) is an
\emph{answer} to $(L,\theta)$.
Else, the derivation is \emph{failed}.
We denote by $\answers{Q}$ the set of answers to a query $Q$.


\section{An Abstract Approach to Modular Logic Programs}

There have been several proposals to date for supporting modularity in
logic programs, all of which are based on performing a partition of
the set of program symbols into modules.
As mentioned before, the two most widely adopted approaches are
referred to as \emph{predicate-based} and \emph{atom-based} module
systems.
In predicate-based module systems all symbols involved in
terms are global, i.e., they belong to a single global \kbd{user}
module --a special module from which all modules import the symbols
and to which all modules can add symbols.
In atom-based module systems~\cite{xsb-journal-2012}
only constants and explicitly exported symbols are global,
while the rest of the symbols are local to their modules.
Ciao~\cite{ciao-modules-cl2000-short} adopts a hybrid approach
which is as in predicate-based systems but
with the possibility of marking a selected set of symbols as local (we
will use this model in the examples in Sec.~\ref{sec:hiding}).
Despite the differences among these module systems,
by performing module resolution applying the appropriate visibility
rules, programs are reducible in all systems to a form that can be
interpreted using the same Prolog-style semantics.
We will use this property in order to abstract our results away from
particular module systems and their symbol visibility rules. To this
end we present a formalization of the ``flattened'' version of a
modular program, where visibility is explicit and is thus independent
of the visibility conventions of specific module systems.
Let $\MS$ denote the set of all \emph{module symbols}.
The \emph{flattened}
form of a modular definite program is
defined as follows:

\begin{definition}[Modular Program]
  \label{def:modprog}
  A \emph{modular program} is a pair
  $( P, \mods{\cdot})$, where
  $P$ is a definite program and $\mods{\cdot}$ is a mapping that assigns
  for each symbol $f \in \FS$ a unique module symbol $m \in \MS$.
  Let $C$ be a clause $H \leftarrow B$ in $P$, $\mods{C} \triangleq \mods{H}$.
  Let $A$ be an atom\footnote{In practice constraints are also located
    in modules. It is trivial to extend the formalization
    to include this, we do not write it explicitly for
    simplicity. }
  or a term of the form $f(\ldots)$. Then $\mods{A} \triangleq \mods{f}$.
\end{definition}

\finalcompression{-1mm}
The $\mods{\cdot}$ mapping
creates a partition
of the clauses in the definite program $P$. We refer to each resulting
equivalence class as a module, and represent it with the module
symbol shared by all clauses in that class.
The set of all symbols defined by a module $m$ is
$\defs{m} = \{ f | f \in \FS, \mods{f} = m, m \in \MS\}$.

\begin{definition}[Interface of a Module]
  \label{def:modprogitf}
  The \emph{interface} of a module $m$ is given by the disjoint sets
  $\exps{m}$ and $\imps{m}$, s.t.
  $\exps{m} \subseteq \defs{m}$ is the subset of the symbols
  defined in $m$ that can appear in other modules, referred to as
  the \emph{export list} of $m$,
  and %
  $\imps{m} = \{ f | f\!\in\!\FS, f \textrm{ is in symbols of }
  \cls{p}, p\!\in\!\defs{m}\} \setminus \defs{m}$ is a superset of
  symbols in the bodies of the predicates of $m$, that are not
  defined in $m$, referred to as the \emph{import list} of $m$.
\end{definition}

To track calls across module boundaries we introduce the notion of
\emph{clause end literal}, a marker of the form $\ret{H}$, where $H$
stands for the head of the parent clause,
as given in the following definition:

  \begin{definition}[Operational Semantics of Modular Programs]
  \label{def:oper-sem-mod}
    We redefine the derivation semantics such that goal sequences are
    of the form $(L,m)::G$ where $L$ is a literal, and $m$ is the
    module from which $L$ was introduced, as shown below.
    Then, a state $S = \state{(L,m)::G}{\theta}$ can be reduced to a state
    $S'$ as follows:
        \finalcompression{-1mm}
    \begin{shortenumerate}
    \item $\reduction{\state{(L,m)::G}{\theta}}{\state{G}{\theta \land L}}$
          if $L$ is a constraint and $\theta \land L$ is satisfiable.
    \item $\reduction{\state{(L,m)::G}{\theta}}%
                     {\state{(B_1,n)::\ldots::(B_k,n)::(\ret{L},n)::G}{\theta}}$
          if $L$ is an atom and
          $\exists (L \leftarrow B_1, \ldots, B_k) \in \cls{L}$
          where $\mods{L}=n$ and it holds that
          $(L\!\in\!\defs{n} \land n\!=\!m)$%
          ~$\bigvee$~%
          $(L\!\in\!\exps{n} \land L\!\in\!\imps{m} \land n\!\neq\!m)$.%
          \label{item:semmodtwo}
    \item $\reduction{\state{(L,m)::G}{\theta}}%
                     {\state{G}{\theta}}$
          if $L$ is a clause return literal $\ret{\_}$.
    \end{shortenumerate}
  \end{definition}

  Basically, for reduction step~\ref{item:semmodtwo} to succeed, the
  $L$ literal should either be defined in module $m$ (and then
  $n = m$) or it should belong to the export list of module $n$ and be
  in the import list of module $m$.


\section{Run-Time Checking of Modular Programs}
\label{mod-rtchecks}

\paragraph{Assertion Language}
We assume that program specifications are provided by means of
assertions: linguistic constructions that allow expressing properties
of programs.
For concreteness we will use the \kbd{pred} assertions of the Ciao
assertion language~\cite{%
  prog-glob-an-shorter,%
  assrt-theoret-framework-lopstr99-shorter,hermenegildo11:ciao-design-tplp-shorter},
following the formalization
of~\cite{asrHO-ppdp2014-shorter,optchk-ppdp2016-shorter}.
Such \kbd{pred} assertions
define the set of all admissible preconditions for
a given predicate, and for each such pre-condition,
a corresponding post-condition.
These pre- and post-conditions are formulas containing literals
defined by predicates that are specially labeled as
\emph{properties}.
Properties and the other
predicates composing the program are written in the same language.
This approach is motivated by the direct correspondence between the
declarative and operational semantics of constraint logic programs.
In what follows we refer to these literals corresponding to properties
as \emph{prop} literals.
The predicate symbols of \emph{prop} literals are module-qualified in
the same way as those of the other program literals.
\begin{example}[Property]
  The following property
  describes a sorted list:
\finalcompression{-2mm}
\begin{verbatim}
sorted([]).  sorted([_]).  sorted([X,Y|L]) :- X =< Y, sorted([Y|L]).
\end{verbatim}
\finalcompression{-2mm}
  i.e., $\llbracket sorted(A) \rrbracket =$
 {\small $\{A=[], A=[B], A=[B,C|D] \wedge B\leq C \wedge E = [C|D] \wedge sorted(E)\}.$}
\end{example}

The left part of Fig.~\ref{fig:asr-and-asrconds} shows a set of
assertions for a predicate (identified by a normalized atom $Head$).
The $Pre_i$ and $Post_i$ fields are conjunctions%
\footnote{In the general case $Pre$ and $Post$ can be DNF formulas of
  \emph{prop} literals but we limit them to conjunctions herein for
  simplicity of presentation.}  of \emph{prop} literals that refer to
the variables of $Head$.
Informally, such a set of assertions states that in any execution state
$\state{(Head,m) :: G}{\theta}$ at least one of the $Pre_i$ conditions
should hold, and that, given the $(Pre_i,Post_i)$ pair(s) where
$Pre_i$ holds, then, if the predicate succeeds, the corresponding
$Post_i$ should hold upon success.
\begin{figure}[t]
\begin{minipage}{0.45\textwidth}
\begin{small}
  \[
  \begin{array}{l}
    \kbd{:-}\kbd{ pred } Head \kbd{ : } Pre_1 \kbd{ => } Post_1 \kbd{.}
  \\
    \ldots
  \\
    \kbd{:-}\kbd{ pred } Head \kbd{ : } Pre_n \kbd{ => } Post_n \kbd{.}
  \end{array}
  \]
\end{small}
\end{minipage}
\begin{minipage}{0.45\textwidth}
\begin{small}
  \[
    C_i = \left\{
    \begin{array}{ll}
      c_i.\callsAsr{Head}{\bigvee _{j = 1}^{n} Pre_j}
    & i = 0
    \\
      c_i.\successAsr{Head}{Pre_i}{Post_i}
    & i = 1..n
    \end{array}
    \right.
  \]
\end{small}
\end{minipage}
\finalcompression{-5mm}

\caption{Correspondence between assertions and assertion conditions.}
\label{fig:asr-and-asrconds}

\end{figure}
We denote the set of assertions
for a predicate represented by
$Head$
by $\A(Head)$, and the set of all assertions in a program by $\A$.

In our formalization, rather than using the assertions for a predicate
directly, we use instead a normalized form which we refer to as the
set of \emph{assertion conditions} for that predicate, denoted as
$\AC(Head) = \{ C_0, C_1, \ldots , C_n\}$, as shown in
Fig.~\ref{fig:asr-and-asrconds}, right. The $c_i$ are identifiers
which are unique for each assertion condition.
The $\callsAsr{Head}{\ldots}$ conditions encode the check
that ensures that the calls to the predicate represented by the $Head$
literal are within those admissible by the set of assertions, and we
thus call them the \emph{calls assertion conditions}.
The $\successAsr{Head}{Pre_i}{Post_i}$ conditions encode
the checks for compliance of the successes for particular sets of calls,
and we thus call them the \emph{success assertion conditions}.
If there are no assertions associated with $Head$ then the corresponding
set of assertion conditions is empty.
The set of assertion conditions for a program, denoted $\AC$ is the
union of the assertion conditions for each of the predicates in the
program, and is derived from the 
set $\A$ of all assertions in the program.

\paragraph{Semantics with Run-time Checking of Assertions and Modules}
We now present the operational semantics with assertions for modular
programs, which checks whether assertion conditions hold or not while
computing the derivations from a query in a modular program.
The identifiers of the assertion conditions (the $c_i$) are used to
keep track of any violated assertion conditions.
The $\errchk{}$ literal denotes a special goal that marks a derivation
finished because of the violation of the assertion condition with
identifier $\asrId{}$.
A finished derivation from a query $(L,\theta)$ is now
\emph{successful} if the last state is of the form
$\state{\emptyGoal}{\theta'}$,
\emph{erroneous} if the last state is of the form
$\state{\errchk{}}{\theta'}$, or
\emph{failed} otherwise.
The set of derivations for a program from its set of queries \Q\
using the semantics with run-time checking of assertions is denoted
by $\derivationsRtc{\Q}$.
We also extend the clause return literal to the form
$\retchk{H}{\mathcal{C}}$, where $\mathcal{C}$ is the set of
identifiers $\asrId{i}$ of the assertion conditions that should be
checked at that derivation point.
A literal $L$ \emph{succeeds trivially} for $\theta$ in program
$P$, denoted $\theta \Rightarrow_P L$, iff $\exists \theta'\in
\answers{(L,\theta)}$ such that $\theta\models\theta'$.
Intuitively, a literal $L$ succeeds trivially if $L$ succeeds for $\theta$
without adding new constraints to $\theta$.
This notion captures the checking of properties and we will thus often
refer to this operation as ``checking $L$ in the context of
$\theta$.'' \footnote{Note that even if several assertion conditions
  may be violated at the same time, we consider only the first one of
  them. The ordering is only imposed by the implementation and does
  not affect the semantics.}

\begin{definition}[Operational Semantics for Modular Programs with
Run-time Checking]
A state $S = \state{(L,m)::G}{\theta}$ 
can be
\emph{reduced} to a state $S'$,
denoted \reductionRtc{S}{S'},
as follows:

\finalcompression{-2mm}
\begin{shortenumerate}
  \item If $L$ is a constraint then 
    $S'\!=\!\state{G}{\theta \land L}$
    if $\theta \land L$ is satisfiable.
  \item
    \label{item:semassrttwo}
    If $L$ is an atom and
    $\exists (L \leftarrow B_1, \ldots, B_k) \in \cls{L}$,
    then the new state is 
    \finalcompression{-2mm}
          \[
        S' = \left\{
        \begin{array}{l}
          \state{\errchk{}}{\theta}
          \textrm{~~~~~~~~if } \exists \;
          \labCallsAsr{}{L}{Pre} \in \AC(L)
          ~\land~ \theta \not\Rightarrow_P Pre 
        \\
          \state{(B_1,n)::\ldots::(B_k,n)::(\retchk{L}{\mathcal{C}},n)::G}{\theta}
          \textrm{~~~~otherwise}
        \end{array}
        \right.
     \]

     \finalcompression{-2mm}
          s.t.\
     $\mathcal{C} = \{ \asrId{i} ~|~ \labSuccessAsr{i}{L}{Pre_i}{Post_i}
     \in \AC(L) \;\land\; \theta \Rightarrow_P Pre_i \}$
     where $\mods{L}=n$ and it holds that
          $(L\!\in\!\defs{n} \land n\!=\!m)$%
          ~$\bigvee$~%
          $(L\!\in\!\exps{n} \land L\!\in\!\imps{m} \land n\!\neq\!m)$
  \item
    \label{item:semassrtthree}
    If $L$ is a clause return literal $\retchk{\_}{\mathcal{C}}$, then
    \finalcompression{-2mm}
        \[
      S' = \left\{
      \begin{array}{ll}
        \state{\errchk{}}{\theta}
      &
        \textrm{if}~ \exists\; \asrId{} \in \mathcal{C}
        \textrm{~s.t.~} \labSuccessAsr{}{L'}{\_}{Post} \in \AC(L')
        ~\land~ \theta \not\Rightarrow_P Post 
      \\
         \state{G}{\theta}
      &
        \textrm{otherwise}
      \end{array}
      \right.
    \]
  \end{shortenumerate}
\end{definition}

Theorem~\ref{thm:rtc-correct} below on the correctness of the
operational semantics with run-time checking can be straightforwardly
adapted from~\cite{asrHO-ppdp2014-shorter}.\footnote{The formal
  definition of the equivalence relation on derivations, as well as
  proofs for the theorems and lemmas can be found in
  Appendix~\ref{app:proofs}.}
The completeness of this operational semantics as presented in
Theorem~\ref{thm:rtc-complete} below can only be proved for
\emph{partial} program derivations, as the new semantics introduces
the $\err{}$ literal that directly replaces the goal sequence of a
state in which a violation of an assertion condition
occurs.

\begin{theorem}[Correctness Under Assertion Checking]
  \label{thm:rtc-correct}
  For any tuple $(P,\Q,\A)$
  it holds that
  $ \forall D'\!\in\!\derivationsRtc{\Q}~\exists D\!\in\!\derivations{\Q}
  \text{~s.t.~} D'\, \text{~is equivalent to~} \,D$
  (including partial derivations).
\end{theorem}

\begin{theorem}[Partial Completeness Under Assertion Checking]
  \label{thm:rtc-complete}
  For any tuple $(P,\Q,\A)$
  it holds that
   $ \forall D = (S_1,\ldots,S_k,S_{k+1},\ldots,S_n)\!\in\!\derivations{\Q}
    ~\exists D'\!\in\!\derivationsRtc{\Q}
   \text{~s.t.~} D'$ is equivalent to $D$ or $(S_1,\ldots,S_k,\state{\errchk{}}{\_})$.
\end{theorem}


\section{Shallow Run-Time Checking}
\label{sec:hiding}

As mentioned before, the main advantage of modular programming is that
it allows safe local reasoning on modules, since two different modules
are not allowed to contribute clauses to the same
predicate.\footnote{In practice, an exception
  are
  \kbd{multifile} predicates. However, since
  they need to
  be declared explicitly, local reasoning is still valid assuming
  conservative semantics (e.g., \emph{topmost} abstract values) for
  them.}
Our purpose herein is to study how in systems where the visibility of
function symbols can be controlled, similar reasoning can be performed
at the level of terms, and in particular how such reasoning can be
applied to reducing the overhead of run-time checks. %
We will refer to these reduced checks as \emph{shallow} run-time
checks, which we will formally define later in this section.
We start by recalling how in cases where the visibility of terms
function symbols can be controlled, this reasoning is impossible without
global (inter-modular) program analysis,
using the following example:

\begin{myexample}
\label{ex:nohidden}
Consider a module $m1$ that exports a single predicate \texttt{p/1} that
constructs \texttt{point/1} terms:
\finalcompression{-1mm}
\begin{small}
\begin{verbatim}
:- module(m1, [p/1, r/0]).    % m1 declared, p/1 and r/0 are exported
p(A) :- A = point(B), B = 1.  % A = user:point(1)
:- use_module(m2,[q/1]).      % import q/1 from a module m2
r :- X = point(2), q(X).      % X = user:point(2)
\end{verbatim}
\end{small}
\finalcompression{-1mm}
Here, we want to reason about the terms that can appear during
program execution at several specific program points:
(a) before we call \kbd{p/1}
(point at which execution enters module \kbd{m1});
(b) when the call to \kbd{p/1} succeeds
(point at which execution leaves the module);
and (c) before we call \kbd{q/1}
(point at which execution enters another module).
When we exit the module at points (b) and (c)
we know that in any \kbd{point(X)} constructed in $m1$ either $X = 1$
or $X = 2$.
However, when we enter module $m1$ at point (a)
\kbd{A} could have been bound by the calling module to any term
including, e.g.,
\kbd{point([4,2])}, \kbd{point(2)}, \kbd{point(a)}, \kbd{point(1)}, etc.,
since the use of the \kbd{point/1} functor is not restricted.
\end{myexample}

Now we will consider the case where the visibility of terms can be
controlled. We start by defining the following notion:

\begin{definition}[Hidden Functors of a Module]
The set of hidden functors of a module is the set of functors that appear
in the module that are local and non-exported.

\end{definition}

\finalcompression{-2mm}
\begin{myexample}
  In this example we mark instead the \kbd{point/1} symbol as
  hidden. We use Ciao module system
  notation~\cite{ciao-modules-cl2000-short}, where all function
  symbols belong to \kbd{user}, unless marked with a \kbd{:- hide f/N}
  declaration. Such symbols are hidden, i.e., local and not exported.%
  \footnote{Note that this can be achieved in other systems: e.g., in
    XSB~\cite{xsb-journal-2012} it can be done with a \kbd{:- local/1}
    declaration, combined with not exporting the symbol.
  }

\finalcompression{-3mm}
\begin{small}
\begin{verbatim}
:- module(m1, [p/1, r/0]).
:- hide point/1.              % point/0 is restricted to m1
p(A) :- A = point(B), B = 1.  % m1:point(1), not user:point(1)
:- use_module(m2,[q/1]).
r :- X = point(2), q(X).      % m1:point(2) escapes through call to q/1
\end{verbatim}
\end{small}
\finalcompression{-2mm}

Let us consider the same program points as in Example~\ref{ex:nohidden}.
When we exit the module, we can infer the same results, but with
\kbd{m1:point/1} instead of \kbd{user:point/1}.
Now, if we see the \kbd{m1:point(X)} term at point (a) we know that
it has been constructed in $m1$, and
the \kbd{X} has to be bound to either $1$ or $2$, because the code
that can create bindings for $X$ is only located in $m1$ (and the
\kbd{point/1} terms are passed outside the module at points (b) and
(c)).
\end{myexample}

As mentioned before, these considerations will allow us to use an
optimized form of checking that we refer to as \emph{shallow
  checking}. In order to formalize this notion,
we start by defining all possible
terms that may exist outside
a module $m$ as its \emph{escaping terms}. We will also introduce the
notion of \emph{shallow properties} as the specialization of the
definition of these properties w.r.t.\
these escaping terms, and we will present algorithms to compute such
shallow versions of properties.

\begin{definition}[Visible Terms at a State]
  \label{def:vis-terms}
  A property that represents all terms that are visible in a state
  $S = \state{(L,\_)::G}{\theta}$ of some derivation
  $D \in \derivationsRtc{\Q}$ for a tuple $(P,\Q,\A)$ is
  $\vis{S}{X} \equiv \bigvee_{V \in \vars{L}} ( X\!=\!V \land \theta)$
  where $\vars{L}$ denotes the set of variables of literal $L$.
\end{definition}

\begin{definition}[Escaping terms]
  \label{def:esc-terms}
  Consider all states $S$ in all derivations $D \in \derivationsRtc{\Q}$
  of any tuple $(P,\Q,\A)$
  where $P$ imports a given module $m$.
  A property that represents escaping terms
  w.r.t.\ $m$ is 
  $\esc{m}{X} \equiv \bigvee \vis{S}{X}$
  for each
  $S = \state{(\_,n)::\_}{\_}$ with $n \neq m$.
\end{definition}

The set of all public symbols to which a variable $X$ can be bound
is denoted as $\usr{X} = \{ X | \mods{X} = \kbd{user} \}$.
The following lemma states that it is enough to consider the states at
the module boundaries to compute $\esc{m}{X}$:

\begin{lemma}[Escaping at the Boundaries]
  \label{lemma:esc-boundaries}
  Consider all derivation steps $\reductionRtc{S_1}{S_2}$
  where $S_1 = \state{(L_1,m)::\_}{\_}$
  and $S_2 = \state{(L_2,n)::\_}{\theta}$
  with $n \neq m$.
  That is, the derivation steps when calling a predicate at $n$ from
  $m$ (if $L_1$ is a literal) or when returning from $m$ to module $n$
  (if $L_1$ is $\ret{\_}$).
  Let $\esc{m'}{X}$ be the \emph{smallest property} (i.e., the
  property with the smallest model) such that
  $\theta \Rightarrow_{P} \esc{m'}{X}$ for each variable $X$ in
  the literal $L_2$, and $\usr{X} \Rightarrow_{P} \esc{m'}{X}$.
  Then $\esc{m'}{X} \lor \usr{X}$ is equivalent to
  $\esc{m}{X}$.
\end{lemma}

\begin{algorithm}[t]
  \caption{\ \ \ \textsc{Escaping\_Terms}\label{fig:esc-terms}}
  \begin{algorithmic}[1]
    \Function{\sc Escaping\_Terms}{$M$}
    \State $\textit{Def}$ := $\usr{X}$
    \ForAll{$L$ exported from $M$}
    \ForAll{$\labSuccessAsr{}{L}{\_}{Post} \in \AC(L)$}
    \ForAll{$P \in \textsc{LitNames}(Post, \textit{vars}(L)$)}
    \State $\textit{Def}$ := $\textit{Def} \sqcup P(X)$
    \EndFor
    \EndFor
    \EndFor
    \ForAll{$L$ imported from $M$}
    \ForAll{$\labCallsAsr{}{L}{Pre} \in \AC(L)$}
    \ForAll{$P \in \textsc{LitNames}(Pre, \textit{vars}(L)$)}
    \State $\textit{Def}$ := $\textit{Def} \sqcup P(X)$
    \EndFor
    \EndFor
    \EndFor
    \State \Return ($\esc{m}{X} \leftarrow \textit{Def}$)
    \EndFunction

    \Function{\sc LitNames}{$G, Args$}
    \State \Return set of $P$ such that $A \in Args$ and $G = (\ldots \wedge P(A) \wedge \ldots)$
    \EndFunction
  \end{algorithmic}
\end{algorithm}

Algorithm~\ref{fig:esc-terms} computes an
over-approximation of the $\esc{m}{X}$ property. The algorithm
has two parts. First, it loops over the exported predicates in module
$m$. For each exported predicate we use $Post$ from the success
assertion conditions as a safe over-approximation of the constraints
that can be introduced during the execution of the predicate. We
compute the union ($\sqcup$, which is equivalent to $\vee$ but it can
sometimes simplify the representation) of all properties that restrict
any variable argument in $Post$.
The second part of the algorithm performs the same operation on all
the properties specified in the
$Pre$ of the calls assertions conditions. This is a safe approximation
of the constraints that can be \emph{leaked} to other modules called
from $m$.

Note that the algorithm
can
use analysis information to detect more precise calls to the imported
predicates, as well as more precise successes of the exported
predicates, than those specified in the assertion conditions present
in the program.

\begin{lemma}[Correctness of \textsc{Escaping\_Terms}]
  \label{lemma:escape-correct}
  The \textsc{Escaping\_Terms} algorithm computes a safe
  (over)approximation to $\esc{m}{X}$ (when using the
  operational semantics with assertions).
\end{lemma}

\paragraph{Shallow Properties}
Shallow run-time checking consists in using \emph{shallow} versions of
properties in the run-time checks for the calls across module
boundaries.
While this notion could be added directly to the operational
semantics, we will present it as a program transformation based on the
generation of shallow versions of the properties, since this also
provides a direct implementation path.

\begin{myexample}
  Assume that the set of escaping terms of $m$ contains \texttt{point(1)} and
  it does not contain the more general \texttt{point(\_)}.
  Consider the property:\\ \texttt{intpoint(point(X))~:-~int(X)}.
  Checking 
  \texttt{intpoint(A)} at any program
  point outside $m$ must check first that \texttt{A} is instantiated
  to \texttt{point(X)} and that \texttt{X} is instantiated to an
  integer (\texttt{int(X)}). However, the escaping terms show that it
  is not possible for a variable to be bound to \texttt{point(X)}
  without \texttt{X=1}. Thus, the latter check is redundant.
  We can compute the optimized -- or \emph{shallow} --
  version of \texttt{intpoint/1} in the context of all execution
  points external to $m$ as \texttt{intpoint(point(\_))}.
\end{myexample}

\medskip
Let $\textsc{Spec}(L,Pre)$ generate a
specialized version $L'$ of
predicate $L$ w.r.t.\ the calls given by $Pre$
(see~\cite{ai-with-specs-sas06-short}). It holds that for all
$\theta$, $\theta \Rightarrow_{P} L$ iff $\theta \wedge Pre
\Rightarrow_{P} L'$.

\begin{definition}[Shallow property]
  \label{def:shallow-prop}
  The shallow version of a property $L(X)$ w.r.t. module $m$ is
  denoted as $L(X)^{\#}$, and computed as $\textsc{Spec}(L(X),Q(X))$,
  where $Q(X)$ is a (safe) approximation of the escaping terms of $m$
  ($\textsc{Escaping\_Terms}(m)$).
\end{definition}

\begin{algorithm}[t]
  \caption{\ \ \ \textsc{Shallow\_Interface}\label{fig:shallow-itf}}
  \begin{algorithmic}[1]
    \Function{\sc Shallow\_Interface}{$M$}
    \State Let $M'$ be $M$ with wrappers for exported predicates
    \State ~~(to differentiate internal from external calls)
    \State Let $Q(X) := \textsc{Escaping\_Terms}(M')$
    \ForAll{$L$ exported from $M$}
    \ForAll{$\labCallsAsr{}{L}{Pre} \in \AC(L)$}
    \State Update $\AC(L)$ with $\labCallsAsr{}{L}{Pre^{\#}}$
    \EndFor
    \ForAll{$\labSuccessAsr{}{L}{Pre}{Post} \in \AC(L)$}
    \State Update $\AC(L)$ with $\labSuccessAsr{}{L}{Pre^{\#}}{Post}$
    \EndFor
    \EndFor
    \State \Return $M'$
    \EndFunction
  \end{algorithmic}
\end{algorithm}
Algorithm~\ref{fig:shallow-itf} computes the
optimized version of a module interface using shallow checks.
It first introduces wrappers for the exported predicates,
i.e., predicates \texttt{p(X) :- p'(X)}, 
renaming all internal
occurrences of \texttt{p} by \texttt{p'}.
Then it computes an approximation $Q(X)$ of the escaping terms of $M$.
Finally, it updates all $Pre$ in calls and success assertion
conditions, for all exported predicates, with their shallow version
$Pre^{\#}$.
We compute the shallow version of a conjunction of literals
$Pre=\bigwedge_i L_i$
as $Pre^{\#}=\bigwedge_i L_i^{\#}$.

\begin{theorem}[Correctness of \textsc{Shallow\_Interface}]
  \label{theorem:shallow-op-correct}
  Replacing a module $m$ in a larger program by its shallow version
  does not alter the (run-time checking) operational semantics.
\end{theorem}

\paragraph{Discussion about precision}
The presence of any \emph{top} 
properties in the calls or success assertion conditions will propagate
to the end in the \textsc{Escaping\_Terms} algorithm (see
Algorithm~\ref{fig:esc-terms}).
For a significant class of programs, this is not a problem as long as we
can provide or infer precise assertions which do not use this top element.
Note that $\usr{X}$, since it has a void intersection with
any hidden term, does not represent a problem.
For example, many generic Prolog term manipulation predicates (e.g.,
\texttt{functor/3}) typically accept a \emph{top} element in their
calls conditions. We restrict these predicates to work only on
$\mathsf{user}$ (i.e., not hidden) symbols.\footnote{This can be
  implemented very efficiently with a simple bit check on the atom
  properties and does not impact the execution.}
More sophisticated solutions, that are outside the scope of this
paper, include: producing monolithic libraries (creating versions of
the imported modules and using abstract interpretation to obtain more
precise assertion conditions); or disabling shallow checking (e.g.,
with a dynamic flag) until the execution exits the context of $m$
(which is correct except for the case when terms are dynamically asserted).

\paragraph{Multi-library scenarios}
Recall that properties can be exported and used in assertions
from other modules.
The shallow version of properties in $m$ are safe to be used not only
at the module boundaries but also in any other assertion check outside
$m$.
Computing the shallow optimization can be performed per-library,
without strictly requiring intermodular analysis. However, in some cases
intermodular analysis may improve the precision of escaping terms and
allow more aggressive optimizations.


\section{Experimental Results}

We
explore the effectiveness of the combination of term hiding and
shallow checking in the reusable library context, i.e., in libraries that
use (some) hidden terms in their data structures and offer an
interface for clients to access/manipulate such terms.
We study
the four assertion checking modes of~\cite{optchk-ppdp2016-shorter}:
\emph{Unsafe} (no library assertions are checked),
\emph{Client-Safe} (checks are generated only for the assertions
of the predicates exported by the library, assertions for the internal
library predicates are not checked),
\emph{Safe-RT} (checks are generated from assertions both for internal
 and exported library predicates),
and \emph{Safe-CT+RT} (like \emph{RT}, but analysis information is used
to clear as many checks as possible at compile-time).
We use the lightweight instrumentation scheme
from~\cite{cached-rtchecks-iclp2015-shorter} for generating the run-time checks
from the program assertions.
For eliminating the run-time checks via static analysis
we reuse the Ciao verification framework, including the extensions
from~\cite{optchk-ppdp2016-shorter}.
We concentrate in
these experiments on shape analysis (regular types).


In our experiments each benchmark is composed of a library and a
client/driver.
We have selected a set of Prolog libraries that implement tree-based
data structures.
Libraries \kbd{B-tree} and
\kbd{binary} \kbd{tree} were taken from the Ciao
sources; libraries \kbd{AVL-tree}, \kbd{RB-tree}, and \kbd{heap} were
adapted from
YAP,
adding similar assertions to those of the Ciao libraries.
Table~\ref{tbl:bmk-metrics} shows some statistics for these libraries:
number of lines of code (LOC), size of the object file (Size KB),
the number of assertions in the library specification considered
(Pred Assertions), and the number of hidden functors per library
($\#$ Hidden Symbols).

\begin{table}[t]
  \caption{Benchmark metrics.}
  \label{tbl:bmk-metrics}
  \begin{minipage}{\textwidth}
  \centering
  \begin{tabular}{r|r|r|c|c}
\hline\hline
    Name          & LOC & Size (KB)  & Pred Assertions & $\#$ Hidden Symbols
\\ \hline
\kbd{AVL-tree}     & 147 &  16.7      & 20        & 2 \\
\kbd{B-tree}       & 240 &  22.1      & 18        & 3 \\
\kbd{Binary tree}  &  58 &   8.3      &  6        & 2 \\
\kbd{Heap}         & 139 &  15.1      & 12        & 3 \\
\kbd{RB-tree}      & 678 & 121.8      & 20        & 4 \\

\hline\hline
  \end{tabular}
  \end{minipage}
\end{table}
In order to focus
on the assertions of the library operations used in the benchmarks
(where by an operation we mean the set of predicates implementing it)
we do not count in the tables
the
assertions for
library predicates
not directly involved in those operations.
Library assertions contain
instantiation (moded) regular types.\footnote{A simple example of
  assertions, escaping terms, and shallow checks can be found in
  Appendix~\ref{app:example}. Full
  plots for all benchmarks can be found in Appendix~\ref{app:plots}.}
For each library we have created two drivers (clients) resulting in
two experiments per library.
In the first one the library operation has constant
($O(1)$) time complexity and the respective run-time check has
$O(N)$ time complexity
(e.g., looking up the value stored at the root of a binary tree and
checking on each lookup that the input term is a binary
tree).
Here a major speedup is expected when using \emph{shallow} run-time
checks, since the checking time dominates operation execution time and
the reduction
due to shallow checking should be more noticeable.
In the second one the library operation has non-constant ($O(log(N))$)
complexity and the respective run-time check $O(N)$ complexity
(e.g., inserting an element in a binary tree and checking on each
insertion that the input term is a tree).
Here obviously a smaller speedup is to be expected
with \emph{shallow} checking.
All experiments were run on a MacBook Pro, 
2.6 GHz Intel Core
i5 processor, 8GB RAM, and under the
Mac OS X 10.12.3 operating
system.






\FPset{\TCiaoPP}{1727}

\FPset{\TavlCHKclsf}{2981}
\FPset{\TavlCHKctrt}{4097}
\FPset{\TavlCHKrtsf}{2750}
\FPset{\TavlCHKunsf}{209}
\FPadd{\TavlFULLclsf}{\TavlCHKclsf}{-\TCiaoPP}
\FPadd{\TavlFULLctrt}{\TavlCHKctrt}{-\TCiaoPP}
\FPadd{\TavlFULLrtsf}{\TavlCHKrtsf}{-\TCiaoPP}
\FPeval{\TavlCLSF}{round(\TavlFULLclsf,0)}
\FPeval{\TavlCTRT}{round(\TavlFULLctrt,0)}
\FPeval{\TavlRTSF}{round(\TavlFULLrtsf,0)}
\FPeval{\TavlUNSF}{round(\TavlCHKunsf,0)}

\FPset{\TbCHKclsf}{2825}
\FPset{\TbCHKrtsf}{3255}
\FPset{\TbCHKunsf}{215}
\FPadd{\TbFULLclsf}{\TbCHKclsf}{-\TCiaoPP}
\FPset{\TbCHKctrt}{4406} \FPadd{\TbFULLctrt}{\TbCHKctrt}{-\TCiaoPP}
\FPadd{\TbFULLrtsf}{\TbCHKrtsf}{-\TCiaoPP}
\FPeval{\TbCLSF}{round(\TbFULLclsf,0)}
\FPeval{\TbCTRT}{round(\TbFULLctrt,0)}
\FPeval{\TbRTSF}{round(\TbFULLrtsf,0)}
\FPeval{\TbUNSF}{round(\TbCHKunsf,0)}

\FPset{\TbinCHKclsf}{2586}
\FPset{\TbinCHKctrt}{3200}
\FPset{\TbinCHKrtsf}{2702}
\FPset{\TbinCHKunsf}{188}
\FPadd{\TbinFULLclsf}{\TbinCHKclsf}{-\TCiaoPP}
\FPadd{\TbinFULLctrt}{\TbinCHKctrt}{-\TCiaoPP}
\FPadd{\TbinFULLrtsf}{\TbinCHKrtsf}{-\TCiaoPP}
\FPeval{\TbinCLSF}{round(\TbinFULLclsf,0)}
\FPeval{\TbinCTRT}{round(\TbinFULLctrt,0)}
\FPeval{\TbinRTSF}{round(\TbinFULLrtsf,0)}
\FPeval{\TbinUNSF}{round(\TbinCHKunsf,0)}

\FPset{\TheapCHKclsf}{2783}
\FPset{\TheapCHKctrt}{3453}
\FPset{\TheapCHKrtsf}{2857}
\FPset{\TheapCHKunsf}{208}
\FPadd{\TheapFULLclsf}{\TheapCHKclsf}{-\TCiaoPP}
\FPadd{\TheapFULLctrt}{\TheapCHKctrt}{-\TCiaoPP}
\FPadd{\TheapFULLrtsf}{\TheapCHKrtsf}{-\TCiaoPP}
\FPeval{\TheapCLSF}{round(\TheapFULLclsf,0)}
\FPeval{\TheapCTRT}{round(\TheapFULLctrt,0)}
\FPeval{\TheapRTSF}{round(\TheapFULLrtsf,0)}
\FPeval{\TheapUNSF}{round(\TheapCHKunsf,0)}

\FPset{\TrbCHKclsf}{3410}
\FPset{\TrbCHKctrt}{4613}
\FPset{\TrbCHKrtsf}{3186}
\FPset{\TrbCHKunsf}{322}
\FPadd{\TrbFULLclsf}{\TrbCHKclsf}{-\TCiaoPP}
\FPadd{\TrbFULLctrt}{\TrbCHKctrt}{-\TCiaoPP}
\FPadd{\TrbFULLrtsf}{\TrbCHKrtsf}{-\TCiaoPP}
\FPeval{\TrbCLSF}{round(\TrbFULLclsf,0)}
\FPeval{\TrbCTRT}{round(\TrbFULLctrt,0)}
\FPeval{\TrbRTSF}{round(\TrbFULLrtsf,0)}
\FPeval{\TrbUNSF}{round(\TrbCHKunsf,0)}


\FPdiv{\TavlCLSFdiv}{\TavlCLSF}{\TavlUNSF} \FPeval{\TavlCLSFratio}{round(\TavlCLSFdiv,0)}
\FPdiv{\TavlRTSFdiv}{\TavlRTSF}{\TavlUNSF} \FPeval{\TavlRTSFratio}{round(\TavlRTSFdiv,0)}
\FPdiv{\TavlCTRTdiv}{\TavlCTRT}{\TavlUNSF} \FPeval{\TavlCTRTratio}{round(\TavlCTRTdiv,0)}

\FPdiv{\TbCLSFdiv}{\TbCLSF}{\TbUNSF} \FPeval{\TbCLSFratio}{round(\TbCLSFdiv,0)}
\FPdiv{\TbRTSFdiv}{\TbRTSF}{\TbUNSF} \FPeval{\TbRTSFratio}{round(\TbRTSFdiv,0)}
\FPdiv{\TbCTRTdiv}{\TbCTRT}{\TbUNSF} \FPeval{\TbCTRTratio}{round(\TbCTRTdiv,0)}

\FPdiv{\TbinCLSFdiv}{\TbinCLSF}{\TbinUNSF} \FPeval{\TbinCLSFratio}{round(\TbinCLSFdiv,0)}
\FPdiv{\TbinRTSFdiv}{\TbinRTSF}{\TbinUNSF} \FPeval{\TbinRTSFratio}{round(\TbinRTSFdiv,0)}
\FPdiv{\TbinCTRTdiv}{\TbinCTRT}{\TbinUNSF} \FPeval{\TbinCTRTratio}{round(\TbinCTRTdiv,0)}

\FPdiv{\TheapCLSFdiv}{\TheapCLSF}{\TheapUNSF} \FPeval{\TheapCLSFratio}{round(\TheapCLSFdiv,0)}
\FPdiv{\TheapRTSFdiv}{\TheapRTSF}{\TheapUNSF} \FPeval{\TheapRTSFratio}{round(\TheapRTSFdiv,0)}
\FPdiv{\TheapCTRTdiv}{\TheapCTRT}{\TheapUNSF} \FPeval{\TheapCTRTratio}{round(\TheapCTRTdiv,0)}

\FPdiv{\TrbCLSFdiv}{\TrbCLSF}{\TrbUNSF} \FPeval{\TrbCLSFratio}{round(\TrbCLSFdiv,0)}
\FPdiv{\TrbRTSFdiv}{\TrbRTSF}{\TrbUNSF} \FPeval{\TrbRTSFratio}{round(\TrbRTSFdiv,0)}
\FPdiv{\TrbCTRTdiv}{\TrbCTRT}{\TrbUNSF} \FPeval{\TrbCTRTratio}{round(\TrbCTRTdiv,0)}



\FPset{\TavlPREPshfrMS}{2.374}   \FPeval{\TavlPREPshfr}{round(\TavlPREPshfrMS,0)}
\FPset{\TavlANAshfrMS}{9.689}    \FPeval{\TavlANAshfr}{round(\TavlANAshfrMS,0)}
\FPset{\TavlPREPetermsMS}{2.193} \FPeval{\TavlPREPeterms}{round(\TavlPREPetermsMS,0)}
\FPset{\TavlANAetermsMS}{31.455} \FPeval{\TavlANAeterms}{round(\TavlANAetermsMS,0)}
\FPset{\TavlACHKMS}{59.405}      \FPeval{\TavlACHK}{round(\TavlACHKMS,0)}

\FPset{\TbPREPshfrMS}{3.135}   \FPeval{\TbPREPshfr}{round(\TbPREPshfrMS,0)}
\FPset{\TbANAshfrMS}{9.282}    \FPeval{\TbANAshfr}{round(\TbANAshfrMS,0)}
\FPset{\TbPREPetermsMS}{3.002} \FPeval{\TbPREPeterms}{round(\TbPREPetermsMS,0)}
\FPset{\TbANAetermsMS}{37.907} \FPeval{\TbANAeterms}{round(\TbANAetermsMS,0)}
\FPset{\TbACHKMS}{90.008}      \FPeval{\TbACHK}{round(\TbACHKMS,0)}

\FPset{\TbinPREPshfrMS}{0.726}   \FPeval{\TbinPREPshfr}{round(\TbinPREPshfrMS,0)}
\FPset{\TbinANAshfrMS}{9.036}    \FPeval{\TbinANAshfr}{round(\TbinANAshfrMS,0)}
\FPset{\TbinPREPetermsMS}{0.733} \FPeval{\TbinPREPeterms}{round(\TbinPREPetermsMS,0)}
\FPset{\TbinANAetermsMS}{14.129} \FPeval{\TbinANAeterms}{round(\TbinANAetermsMS,0)}
\FPset{\TbinACHKMS}{33.201}      \FPeval{\TbinACHK}{round(\TbinACHKMS,0)}

\FPset{\TheapPREPshfrMS}{1.884}   \FPeval{\TheapPREPshfr}{round(\TheapPREPshfrMS,0)}
\FPset{\TheapANAshfrMS}{6.626}    \FPeval{\TheapANAshfr}{round(\TheapANAshfrMS,0)}
\FPset{\TheapPREPetermsMS}{1.634} \FPeval{\TheapPREPeterms}{round(\TheapPREPetermsMS,0)}
\FPset{\TheapANAetermsMS}{24.248} \FPeval{\TheapANAeterms}{round(\TheapANAetermsMS,0)}
\FPset{\TheapACHKMS}{71.356}      \FPeval{\TheapACHK}{round(\TheapACHKMS,0)}

\FPset{\TrbPREPshfrMS}{13.321}   \FPeval{\TrbPREPshfr}{round(\TrbPREPshfrMS,0)}
\FPset{\TrbANAshfrMS}{10.98}     \FPeval{\TrbANAshfr}{round(\TrbANAshfrMS,0)}
\FPset{\TrbPREPetermsMS}{14.092} \FPeval{\TrbPREPeterms}{round(\TrbPREPetermsMS,0)}
\FPset{\TrbANAetermsMS}{35.289}  \FPeval{\TrbANAeterms}{round(\TrbANAetermsMS,0)}
\FPset{\TrbACHKMS}{298.112}      \FPeval{\TrbACHK}{round(\TrbACHKMS,0)}


\FPeval{\TavlANAsum}{\TavlPREPshfr+\TavlANAshfr+\TavlPREPeterms+\TavlANAeterms}
\FPeval{\TavlANA}{round(\TavlANAsum,0)}
\FPdiv{\TavlANAratio}{\TavlANAsum}{\TavlFULLctrt}
\FPmul{\TavlANAproc}{\TavlANAratio}{100}
\FPeval{\TavlANArel}{round(\TavlANAproc,0)}
\FPdiv{\TavlACHKratio}{\TavlACHK}{\TavlFULLctrt}
\FPmul{\TavlACHKproc}{\TavlACHKratio}{100}
\FPeval{\TavlACHKrel}{round(\TavlACHKproc,0)}

\FPeval{\TbANAsum}{\TbPREPshfr+\TbANAshfr+\TbPREPeterms+\TbANAeterms}
\FPeval{\TbANA}{round(\TbANAsum,0)}
\FPdiv{\TbANAratio}{\TbANAsum}{\TbFULLctrt}
\FPmul{\TbANAproc}{\TbANAratio}{100}
\FPeval{\TbANArel}{round(\TbANAproc,0)}
\FPdiv{\TbACHKratio}{\TbACHK}{\TbFULLctrt}
\FPmul{\TbACHKproc}{\TbACHKratio}{100}
\FPeval{\TbACHKrel}{round(\TbACHKproc,0)}

\FPeval{\TbinANAsum}{\TbinPREPshfr+\TbinANAshfr+\TbinPREPeterms+\TbinANAeterms}
\FPeval{\TbinANA}{round(\TbinANAsum,0)}
\FPdiv{\TbinANAratio}{\TbinANAsum}{\TbinFULLctrt}
\FPmul{\TbinANAproc}{\TbinANAratio}{100}
\FPeval{\TbinANArel}{round(\TbinANAproc,0)}
\FPdiv{\TbinACHKratio}{\TbinACHK}{\TbinFULLctrt}
\FPmul{\TbinACHKproc}{\TbinACHKratio}{100}
\FPeval{\TbinACHKrel}{round(\TbinACHKproc,0)}

\FPeval{\TheapANAsum}{\TheapPREPshfr+\TheapANAshfr+\TheapPREPeterms+\TheapANAeterms}
\FPeval{\TheapANA}{round(\TheapANAsum,0)}
\FPdiv{\TheapANAratio}{\TheapANAsum}{\TheapFULLctrt}
\FPmul{\TheapANAproc}{\TheapANAratio}{100}
\FPeval{\TheapANArel}{round(\TheapANAproc,0)}
\FPdiv{\TheapACHKratio}{\TheapACHK}{\TheapFULLctrt}
\FPmul{\TheapACHKproc}{\TheapACHKratio}{100}
\FPeval{\TheapACHKrel}{round(\TheapACHKproc,0)}

\FPeval{\TrbANAsum}{\TrbPREPshfr+\TrbANAshfr+\TrbPREPeterms+\TrbANAeterms}
\FPeval{\TrbANA}{round(\TrbANAsum,0)}
\FPdiv{\TrbANAratio}{\TrbANAsum}{\TrbFULLctrt}
\FPmul{\TrbANAproc}{\TrbANAratio}{100}
\FPeval{\TrbANArel}{round(\TrbANAproc,0)}
\FPdiv{\TrbACHKratio}{\TrbACHK}{\TrbFULLctrt}
\FPmul{\TrbACHKproc}{\TrbACHKratio}{100}
\FPeval{\TrbACHKrel}{round(\TrbACHKproc,0)}

\paragraph{Static Analysis} %
Table~\ref{tbl:bmk-ctrt-ana} presents the detailed compile-time
analysis and checking times for the \emph{Safe-CT+RT} mode.
\begin{table}[t]
  \caption{Static analysis and checking time for benchmarks for 
    the \emph{Safe-CT+RT}
    mode.
  }
  \label{tbl:bmk-ctrt-ana}

  \begin{minipage}{\textwidth}
  \centering
  \begin{tabular}{rrrrrcrc}
\hline\hline

  \multirow{ 2}{*}{Benchmark} & \multicolumn{5}{c}{Analysis time, ms}
            & \multicolumn{2}{c}{Assertions} \\

\cline{2-6}\cline{7-8}
            & prep & \kbd{shfr} & prep & \kbd{eterms} & total
            & checking, ms & unchecked \\
\hline

\kbd{AVL-tree}    & \TavlPREPshfr
                  & \TavlANAshfr
                  & \TavlPREPeterms
                  & \TavlANAeterms
                  & \TavlANA~(\TavlANArel \%)
                  & \TavlACHK~(\TavlACHKrel \%)
                  & 2/20
\\
\kbd{B-tree}      & \TbPREPshfr
                  & \TbANAshfr
                  & \TbPREPeterms
                  & \TbANAeterms
                  & \TbANA~(\TbANArel \%)
                  & \TbACHK~(\TbACHKrel \%)
                  & 3/18
\\
\kbd{Binary tree} & \TbinPREPshfr
                  & \TbinANAshfr
                  & \TbinPREPeterms
                  & \TbinANAeterms
                  & \TbinANA~(\TbinANArel \%)
                  & \TbinACHK~(\TbinACHKrel \%)
                  & 2/6~
\\
\kbd{Heap}        & \TheapPREPshfr
                  & \TheapANAshfr
                  & \TheapPREPeterms
                  & \TheapANAeterms
                  & \TheapANA~(\TheapANArel \%)
                  & \TheapACHK~(\TheapACHKrel \%)
                  & 2/12
\\
\kbd{RB-tree}     & \TrbPREPshfr
                  & \TrbANAshfr
                  & \TrbPREPeterms
                  & \TrbANAeterms
                  & \TrbANA~(\TrbANArel \%)
                  & \TrbACHK~(\TrbACHKrel \%)
                  & 3/20
\\
 \hline\hline

  \end{tabular}
  \end{minipage}
\end{table}
\noindent
Numbers in parentheses indicate the percentage of the total
compilation time spent on analysis, which stays reasonably low even in
the most complicated case ($13\%$ for the \kbd{RB-tree} library).
Nevertheless, the analysis was able to discharge most of the
assertions in our benchmarks, leaving always only 2-3 assertions
unchecked (i.e., that will need run-time checks), for the predicates
of the library operations being benchmarked.

\paragraph{Run-time Checking}
After the static preprocessing phase we have divided our libraries
into two groups:
(a) libraries where the only unchecked assertions left are the ones for
the boundary calls (\kbd{AVL-tree}, \kbd{heap}, and \kbd{binary
tree}),\footnote{%
Due to our reusable library scenario the
analysis of the libraries is performed without any knowledge of the
client and thus the library interface checks must always remain.}
and %
(b) libraries with also some unchecked assertions for internal
calls (\kbd{B-tree} and \kbd{RB-tree}).
We present run time plots\footnote{The current measurements depend on
  the C \kbd{getrusage()} function, that on Mac OS has microsecond
  resolution.} %
for one library of each group.
Since the unchecked assertions in the second group correspond to
internal calls of the $O(log(N))$ operation experiment, we only show
here a set of plots of the $O(1)$ operation experiment for one library,
as these plots are very similar across all benchmarks.


\newcommand{\plotwidth}{0.8\textwidth}

\begin{figure}[t]
  \centering
    \includegraphics[width=\plotwidth]{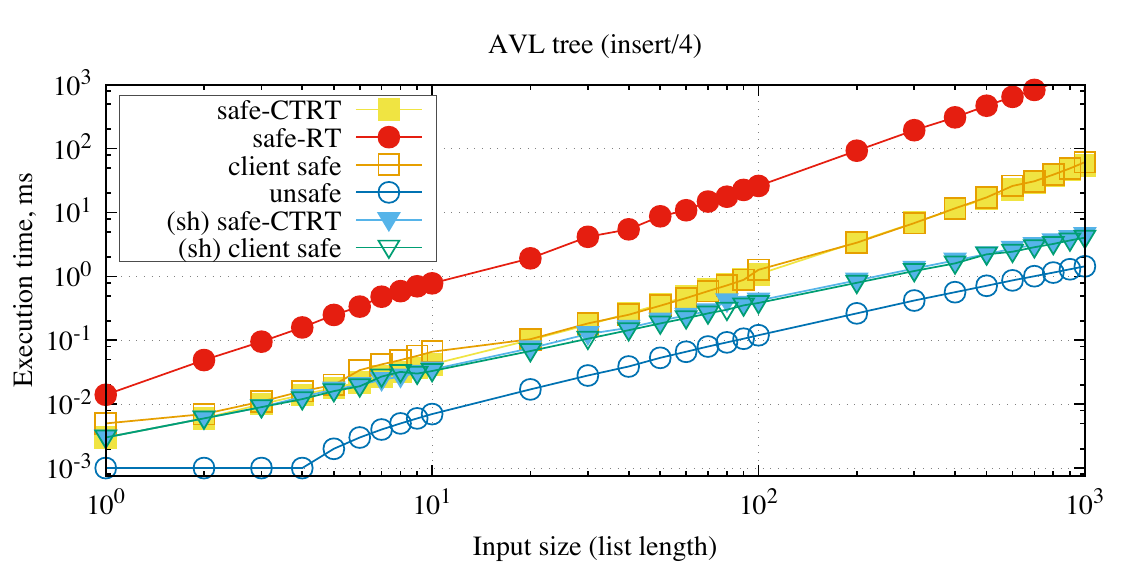}
    \finalcompression{-3mm}
  \caption{Run times in different checking modes,
          \kbd{AVL-tree} library, $O(log(N))$ operation.} 
  \label{fig:plot_b_avl}

\end{figure}

Fig.~\ref{fig:plot_b_avl} illustrates the overhead reductions from using the
shallow
run-time checks in the \kbd{AVL-tree} benchmark for
the $O(N)$ \emph{insert} operation experiment.
This is also the best case that can be achieved for this kind of
operations, since in the \emph{Safe-CT+RT} mode all inner assertions are
discharged statically.
%
\begin{figure}[t]
  \centering
  \includegraphics[width=\plotwidth]{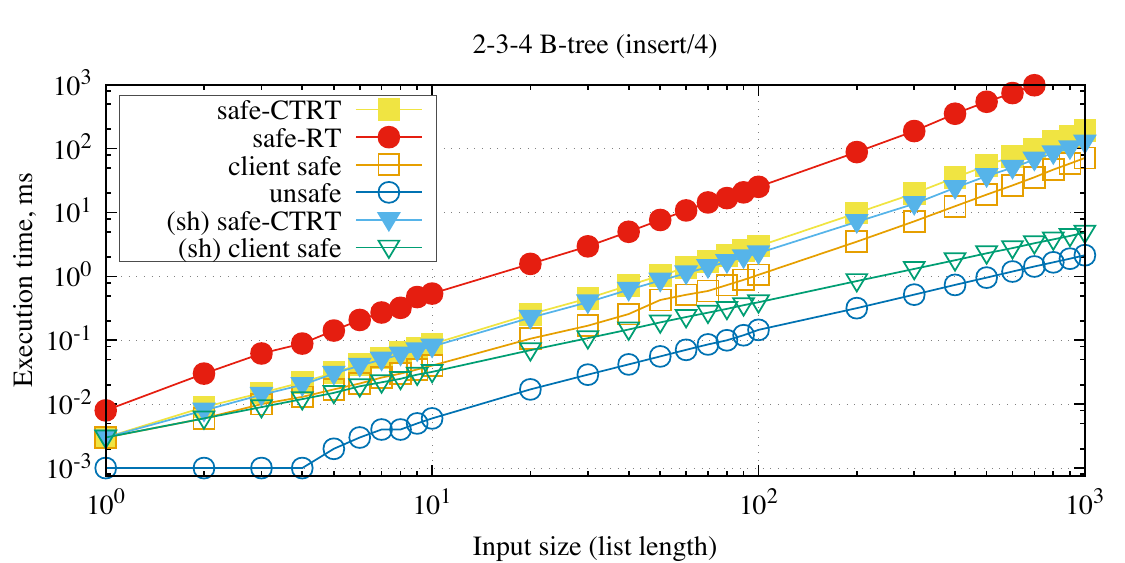}
    \finalcompression{-3mm}
  \caption{Run times in different checking modes,
          \kbd{B-tree} library, $O(log(N))$ operation.}
  \label{fig:plot_b_b}

\end{figure}
%
\begin{figure}[t]
  \centering
  \includegraphics[width=\plotwidth]{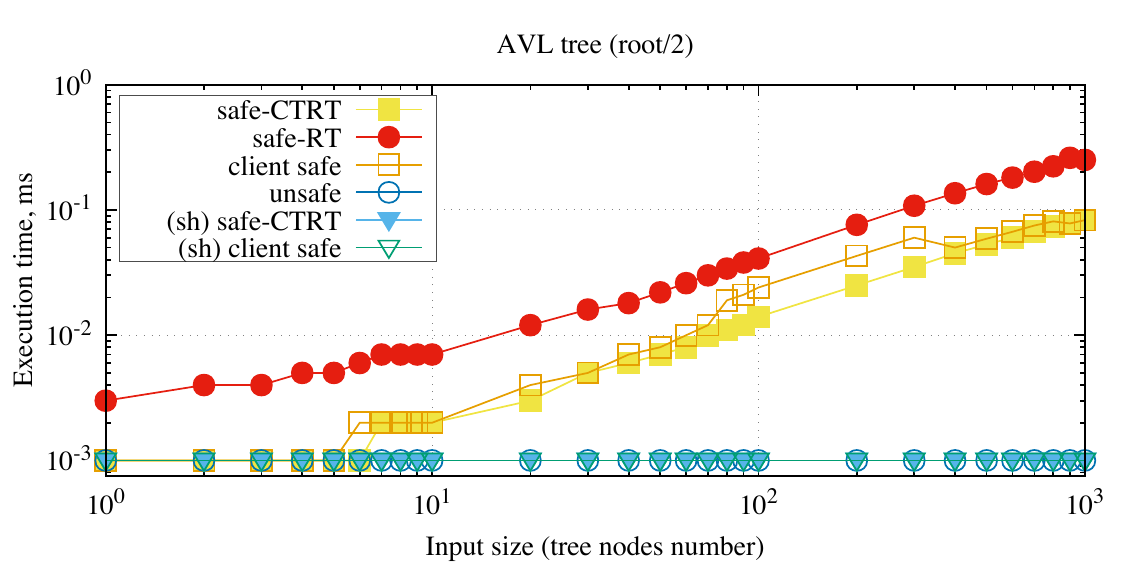}
    \finalcompression{-3mm}
  \caption{Run times in different checking modes, \kbd{AVL-tree}
    library, $O(1)$ operation.}
  \label{fig:plot_b_avl2}

\end{figure}
%
%
Fig.~\ref{fig:plot_b_b} shows the overhead reductions from using the
shallow
checks in the \kbd{B-tree} benchmark for
the $O(log(N))$ \emph{insert} operation experiment.
In contrast with the previous case, here the overhead reductions
achieved by employing shallow checks are dominated by the total check
cost,
and while the overhead reduction is obvious in the \emph{Client-Safe}
mode, it is not significant in the \emph{Safe-CT+RT} mode where some
internal assertion was being checked.
%

Fig.~\ref{fig:plot_b_avl2} presents the overhead reductions in
run-time checking resulting from the use of the shallow
checks in the \kbd{AVL-tree} benchmark for the $O(1)$ \emph{peek}
operation experiment on the root.
As we can see, using shallow checks allows us to obtain constant
overhead on the boundary checks for such cheap operations in all
execution modes but \emph{Safe-RT}.
In summary, the shallow checking technique seems quite effective in
reducing the shape-related run-time checking overheads for the
reusable-library scenario.


\section{Related Work}

\paragraph{Modularity}%
The topic of modules and logic programming has received considerable
attention,
dating back to%
~\cite{WarrenChen87-short,Chen87-short,Miller89} and resulting in
standardization attempts for ISO-Prolog~\cite{standard-modules-final-short}.
Currently, most mature Prolog implementations adopt some flavor of a
module system, \emph{predicate-based} in
SWI~\cite{swi-journal-2012-short},
SICStus~\cite{sicstus-journal-2012-short},
YAP~\cite{yap-journal-2012-short}, and
ECLiPSe~\cite{eclipse-journal-2012-short}, and
\emph{atom-based} in XSB~\cite{xsb-journal-2012}.
As mentioned before,
Ciao~\cite{hermenegildo11:ciao-design-tplp-shorter,ciao-modules-cl2000-short}
uses a hybrid approach, which behaves by default as in predicate-based
systems but
with the possibility of marking a selected set of symbols as hidden,
making it essentially compatible with that of XSB.
Some previous research in the comparative advantages of atom-based
module systems can be found in~\cite{DBLP:conf/iclp/HaemmerleF06-short}.

\paragraph{Parallels with Static Typing and Contracts}%
While traditionally Prolog is untyped, there have been some proposals
for integrating it with type systems, starting
with~\cite{mycr84}.
Several strongly-typed Prolog-based systems have been proposed, notable
examples being Mercury~\cite{mercury-jlp}, G{\"o}del~\cite{goedel-short},
and Visual Prolog~\cite{visual-prolog-short}.
An approach for combining typed and untyped Prolog modules has been
proposed in~\cite{schrijvers08:typed_prolog-short}.
A conceptually similar approach in the world of functional programming
is
\emph{gradual
  typing}~\cite{Siek06gradualtyping,%
DBLP:conf/popl/TakikawaFGNVF16-short}.
The Ciao model offers an (earlier) alternative, closer to \emph{soft
  typing}~\cite{cartwright91:soft_typing-short}, but based on safe
approximations and abstract
interpretation, thus providing a
more general and flexible approach
than
the previous work, since assertions can contain any abstract property
--see~\cite{hermenegildo11:ciao-stop-short} for a discussion of this
topic.
This approach has recently also been applied in a number of contract-based
systems~\cite{clousot-2010-short,DBLP:conf/oopsla/Tobin-HochstadtH12-short,DBLP:conf/pldi/NguyenH15-short},
for which we believe our techniques
can be relevant.

\paragraph{Run-time Checking Optimization}%
High run-time overhead
is a common problem
in systems that
include dynamic
checking~\cite{%
  DBLP:conf/popl/TakikawaFGNVF16-short}.
The impact of global static analysis in reducing run-time checking
overhead has been studied in%
~\cite{optchk-ppdp2016-shorter}.
A complementary approach
is improving the instrumentation of the checks
and combining it with run-time data
caching~\cite{rv2014-short,cached-rtchecks-iclp2015-shorter} or limiting the points
at which 
tests are performed~\cite{testchecks-iclp09-short}.
While these optimizations
can bring significant reductions in overhead, it still remains
dependent on the size of the terms
being checked.
We have shown
herein that even in the challenging context of
calls across open module boundaries it is sometimes possible to
achieve
constant run-time overhead.

\section{Conclusions}

We have described a lightweight modification of a
predicate-based module system to support term hiding and explored the
optimizations that can be achieved with this technique in the context
of combined compile-time/run-time verification.  We have studied the
challenging case of reusable libraries, i.e., library modules that are
pre-compiled independently of the client.
We have shown that with our approach the shape information that can be
inferred
can be enriched significantly
and
large reductions in overhead can be achieved.
The overheads achieved are closer to those of
static
languages, even in the reusable-library context,
without requiring switching to strong typing, which is less natural
in Prolog-style languages, where there is a difference between error
and failure/backtracking.

\vspace{-1mm}
\bibliographystyle{splncs}
\bibliography{%
../../../bibtex/clip/clip.bib,../../../bibtex/clip/general.bib}

\clearpage
\appendix

\section{Main Proofs}
\label{app:proofs}

\subsection{Proof of Theorem~\ref{thm:rtc-correct}}

\begin{definition}[Error-erased Program Derivations]
  \label{def:error-erased-derivation}
  The set of \emph{error-erased} \emph{partial}
  derivations from $\reductionRtc{}{}$ is
  obtained by a syntactic rewriting $\errorErase{(-)}$ that removes
  the error states and sets of assertion condition identifiers from
  the clause end literals.
  It is recursively defined as follows:
  \begin{align*}
    \errorErase{(S_1, \dots, S_m, S_{m+1})}
  &= \left\{
      \begin{array}{lr}
        \errorErase{(S_1, \dots, S_m)}
      &
        \text{if}~ S_{m+1} = \state{\errchk{}}{\_}
      \\
        \errorErase{(S_1, \dots, S_m)} ~\Vert~
        (\errorErase{(S_{m+1})})
      &
        \text{otherwise}
      \end{array}
    \right.
  \\
    \errorErase{\state{G}{\theta}}
  &= \state{\errorErase{G}}{\theta}
  \\
    \errorErase{(L::G)}
  &= \left\{
      \begin{array}{lr}
        (\ret{L'},m)::(\errorErase{G})
                            & \text{~~if}~ L = \retchk{L'}{\_}
      \\
        L::(\errorErase{G}) & \text{~~otherwise}
      \end{array}
    \right.
  \\
    \errorErase{\emptyGoal}
  &= \emptyGoal
  \end{align*}
  \noindent where $\Vert$ stands for sequence concatenation.
  \end{definition}

\begin{proof}[of Theorem~\ref{thm:rtc-correct}]

  We define that $D'\, \text{~is equivalent to~} \,D$ iff
  $\errorErase{(D')} = \,D$.

  Let $D'  = (S'_1,\ldots,S'_k)$,
      $S_i = \state{(L'_i,m_i)}{\theta'_i}$,
      for 
      $Q   = ((L'_1,m_1),\theta'_1) \in {\cal Q}$
      and $\reductionRtc{S'_i}{S'_{i+1}}$.
  Proof by induction on the length $k$ of $D'$:
  \begin{itemize}
  \item Base case ($k=1$). %
    $\errorErase{(S'_1)}
      = \state{\errorErase{(L'_1,m_1)}}{\theta'_1}
      = \state{(L'_1,m_1)}{\theta'_1}
      = S_1$
    since $L'_1$ can be neither the $\retchk{\_}{\_}$ nor $\err{\_}$ literal,
    as they require at least one $\reductionRtc{}{}$ state reduction to
    be reached.
  \item Inductive case (show $k+1$ assuming $k$ holds). %
    In the inductive step it is enough to consider the cases that are
    different in the $\reduction{}{}$ and $\reductionRtc{}{}$
    reductions:
    \begin{itemize}
    \item If $D' = (S'_1,\ldots,S'_k,S'_{k+1})$ and
      $S'_{k+1} = \state{\err{\_}}{\theta'_{k+1}}$ then from
      Def.~\ref{def:error-erased-derivation} it immediately follows
      $\exists D \in \derivations{\Q}$ s.t.
      $\errorErase{(D')}
        = \errorErase{(S'_1,\ldots,S'_k,S'_{k+1})}
        = (S'_1,\ldots,S'_k)
        = D $
    \item If $D' = (S'_1,\ldots,S'_k,S'_{k+1})$ and
      $S'_{k+1} = \state{\retchk{L}{\_}}{\theta'_{k+1}}$ then from
      Def.~\ref{def:error-erased-derivation} it immediately follows
      $\exists D \in \derivations{\Q}$ s.t.
      $\errorErase{(D')}
        = \errorErase{(S'_1,\ldots,S'_k)} \Vert \errorErase{(S'_{k+1})}
        = (S'_1,\ldots,S'_k) \Vert \state{\ret{L}}{\theta'_{k+1}}
        = D$
    \end{itemize}
    \hfill $\qed$
  \end{itemize}
\end{proof}

The proof of Theorem~\ref{thm:rtc-complete} is trivial, based on the
  same reasoning as in the proof of Theorem~\ref{thm:rtc-correct}, and
  is not included for brevity.

\subsection{Proof of Lemma~\ref{lemma:esc-boundaries}}

\begin{proof}
  Let $\esc{m}{X} \equiv \bigvee_i \bigvee_{V \in
    \textit{Vars}_i}(X=V \wedge \theta_i)$ and
  $\esc{m'}{X} \equiv \bigvee_i \bigvee_{V \in
    \textit{Vars}_i'}(X=V \wedge \theta_i')$.
  From the definitions, it can be seen that the set of all
  $\theta_i'$ (at the boundaries, before and after $m$) is a subset
  of all $\theta_i$ (outside $m$).
  The rest of the $\theta_i$ correspond to states not preceded by a
  literal from $m$. For such states $\bigvee_{V \in
    \textit{Vars}_i}(X=V \wedge \theta_i)$ must be:
  1) covered by
  $\usr{X}$ (and thus $\esc{m'}{X}$);
  or 2) contain some $X=f(\ldots)$ with $f$ hidden in $m$. Since $f$
  cannot appear in literals from $n \neq m$ then it must have come
  from some $\theta_b \wedge \theta_o$, where $\theta_b$ is some
  ancestor at the boundaries (already covered), $\theta_o$ is a
  conjunction of constraints introduced outside $m$ (with cannot
  contain $f$), and thus it is more specific and also covered by
  $\esc{m'}{X}$).
\end{proof}

\ \\
\ \\
\subsection{Proof of Lemma~\ref{lemma:escape-correct}}

\begin{proof}
  Let $Q(X) = \textsc{Escaping\_Terms}(m)$, we will show that $Q(X)$
  over-approximates $\esc{m}{X}$.
  Since $\esc{m}{X}$ is equivalent to $\esc{m'}{X}$
  (Lemma~\ref{lemma:esc-boundaries}), it is enough to consider the
  derivation steps at the boundaries. That is, $\reduction{S_1}{S_2}$
  where $S_1 = \state{(L_1,m)::\_}{\_}$
  and $S_2 = \state{(L_2,n)::\_}{\theta}$ with $n \neq m$.
  If $L_1$ is a literal (not $\ret{\_}$) then it corresponds to the
  case of calling an imported predicate. The operational semantics
  ensures that $\theta \Rightarrow_{P} Pre$ and thus $Q(X)$
  over-approximates this case.
  If $L_2$ is $\ret{\_}$ then it corresponds to the case of
  returning from $m$. The operational semantics ensures that $\theta
  \Rightarrow_{P} Post$ and thus $Q(X)$ also over-approximates this
  case.
\end{proof}

\ \\
\ \\
\subsection{Proof of Theorem~\ref{theorem:shallow-op-correct}}

\begin{proof}
  By definition, the transformation only affects the checks for
  $Pre=(\bigwedge_i L_i(X_i))$ conjunctions in assertion conditions of
  exported predicates in $m$.
  These checks correspond to the derivation steps
  $\reduction{S_1}{S_2}$ where
  $S_1 = \state{(\_,n)::G}{\theta}$ and $S_2 = \state{(\_,m)::G}{\_}$
  with $n \neq m$. 
  Let $Q(X)$ be obtained from $\textsc{Escaping\_Terms}(m)$.
  The shallow version $Pre^{\#} = (\bigwedge_i L_i(X_i))^{\#} =
  (\bigwedge_i \textsc{Spec}(L_i(X_i), Q(X_i)))$
  (Definition~\ref{def:shallow-prop}).
  By Definition~\ref{def:esc-terms} it holds that $\theta
  \Rightarrow_{P} (\bigwedge_i \esc{m}{X_i})$.
  By Lemma~\ref{lemma:escape-correct} it holds that $\theta
  \Rightarrow_{P} (\bigwedge_i Q(X_i))$.
  By correctness of $\textsc{Spec}$, since $\theta$ entails each
  $Q(X_i)$, then the full and specialized versions of $L_i$ can be
  interchanged.
\end{proof}

\clearpage

\section{ Example: Computation of Escaping Terms and Shallow Checks \\
          (Code from the \kbd{binary tree} Library)}
\label{app:example}

The code excerpt below below contains
the declarations for hiding locally the \kbd{binary tree} library functors,
the exported \kbd{insert/3} predicate,
its assertion,
and the definitions of the regular types used in this assertion:

{\small%
\begin{verbatim}
:- hide(empty/0).
:- hide(tree/3).

:- regtype val_key/1.
val_key(X) :- int(X).

:- regtype val_tree/1.
val_tree(empty).
val_tree(tree(LC,X,RC)) :- val_tree(LC), val_key(X), val_tree(RC).

:- pred insert(K,T0,T1)  : val_key(K), val_tree(T0), term(T1)
                        => val_key(K), val_tree(T0), val_tree(T1).

insert(X,empty,tree(empty,X,empty)).
insert(X,tree(LC,X,RC),tree(LC,X,RC)).
insert(X,tree(LC,Y,RC),tree(LC_p,Y,RC)) :- X < Y,insert(X,LC,LC_p).
insert(X,tree(LC,Y,RC),tree(LC,Y,RC_p)) :- X > Y,insert(X,RC,RC_p).
\end{verbatim}}

\noindent
The assertion conditions for the \kbd{insert/3} predicate are:
\[
\small
\begin{array}{r}
\labCallsAsr{0}{insert(K,T0,T1)}{val\_key(K), val\_tree(T0), term(T1)}\\
\asrId{1}.\textsf{success}(insert(K,T0,T1),(val\_key(K),val\_tree(T0),term(T1)),\\
(val\_key(K),val\_tree(T0),val\_tree(T1)))
\end{array}
\]

\noindent
%
Denoting the module that contains the library source code as $bt$,
the set of \emph{escaping terms} computed by
Algorithm~\ref{fig:esc-terms} can be represented as the following regular type:


\[
\small
\begin{array}{l}
\kbd{esc}_{bt}\kbd{(}bt\kbd{:empty).}\\
\kbd{esc}_{bt}\kbd{(}bt\kbd{:tree(L,K,R)) :- val\_tree(L), val\_key(K), val\_tree(R).}\\
\kbd{esc}_{bt}\kbd{(X) :- usr(X).}\\
\end{array}
\]

\noindent
where $usr(\_)$ is a property that denotes \kbd{user} terms.
The explicit module qualification $bt\!:$ is used only to clarify that
\kbd{empty/0} and
\kbd{tree/3} are local to module $bt$ and not \kbd{user} functors.
The resulting \emph{shallow interface} produced by
Algorithm~\ref{fig:shallow-itf} is:

{\small%
\begin{verbatim}
:- regtype val_key/1.
val_key(X) :- int(X).

:- regtype val_tree/1.
val_tree(empty).
val_tree(tree(LC,X,RC)) :- val_tree(LC), val_key(X), val_tree(RC).

:- pred insert(K,T0,T1)  : val_key(K), val_tree#(T0), term(T1).
                        => val_key(K), val_tree(T0),  val_tree(T1).
insert(K,T0,T1) :- insert'(K,T0,T1).

:- pred insert'(K,T0,T1)  : val_key(K), val_tree(T0), term(T1)
                         => val_key(K), val_tree(T0), val_tree(T1).
... clauses of insert'/3 ...
\end{verbatim}}

\noindent
where the $\kbd{val\_tree}^{\#}$ property can be materialized as:

{\small%
\begin{verbatim}
val_tree#(empty).
val_tree#(tree(_,_,_)).
\end{verbatim}}

\noindent
The run-time checks instrumentation then can use the shallow
$\kbd{val\_tree}^{\#}\kbd{/1}$ property in the checks for the calls across module
boundaries and the original \kbd{val\_tree/1} property for the calls
inside the $bt$ module.


\clearpage
\section{Additional Plots}
\label{app:plots}
\begin{figure}[h!t]
    \centering
    \includegraphics[width=\textwidth]{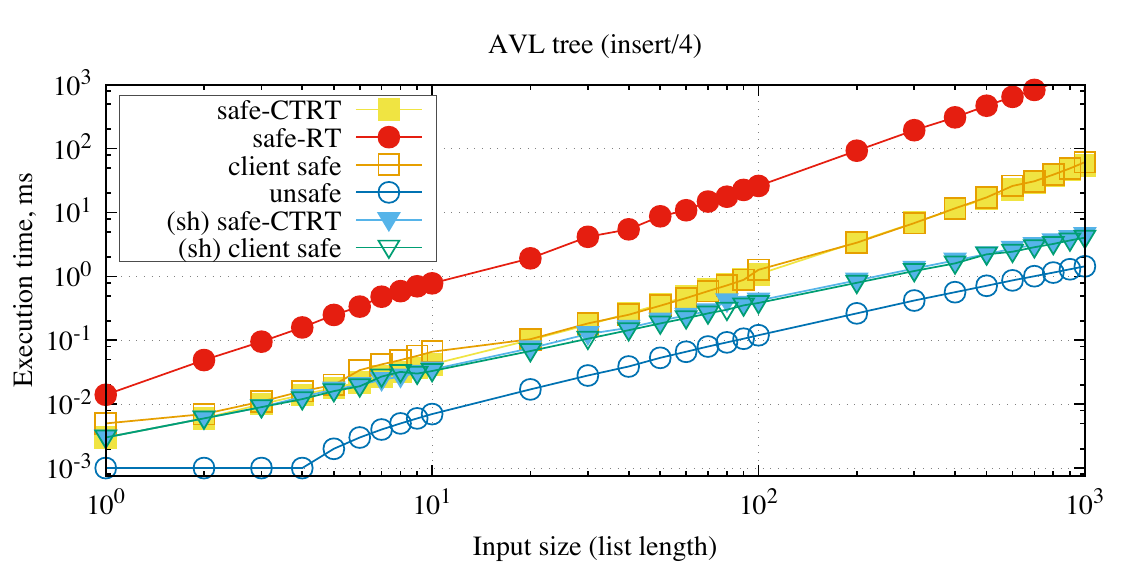}
\caption{Run times for the \kbd{AVL-tree} benchmark in different %
             execution modes, $O(log(N))$ operation + $O(N)$ check complexity}
    \includegraphics[width=\textwidth]{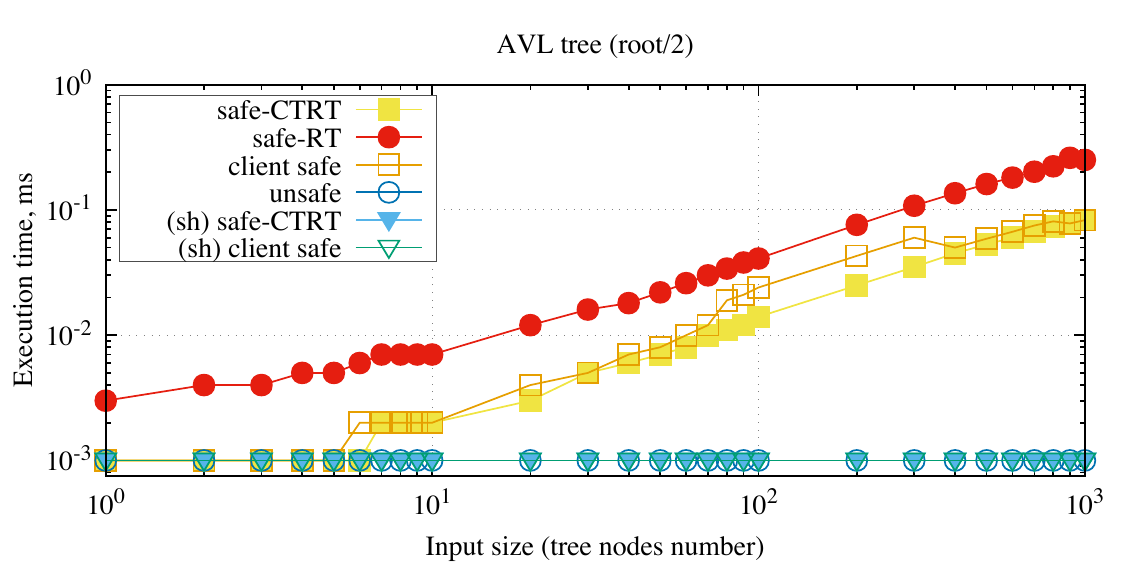}
\caption{Run times for the \kbd{AVL-tree} benchmark in different %
             execution modes, $O(1)$ operation + $O(N)$ check complexity}
\end{figure}

\begin{figure}[t]
    \centering
    \includegraphics[width=\textwidth]{plot_c_b_tree_client.pdf}
\caption{Run times for the \kbd{2-3-4 B-tree} benchmark in different %
             execution modes, $O(log(N))$ operation + $O(N)$ check complexity}
    \includegraphics[width=\textwidth]{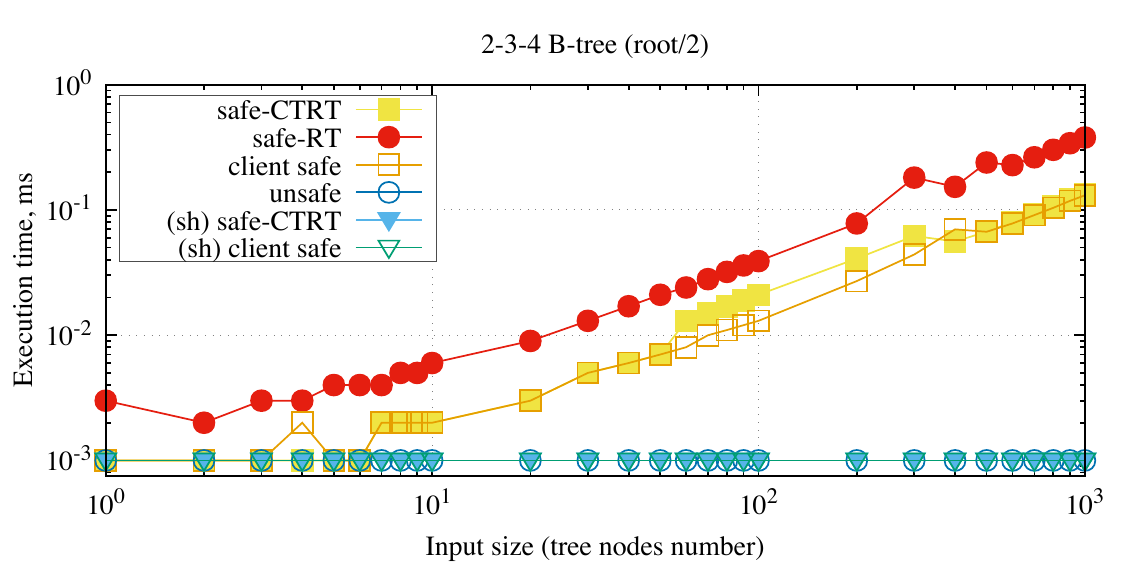}
\caption{Run times for the \kbd{2-3-4 B-tree} benchmark in different %
             execution modes, $O(1)$ operation + $O(N)$ check complexity}
\end{figure}

\begin{figure}[t]
    \centering
    \includegraphics[width=\textwidth]{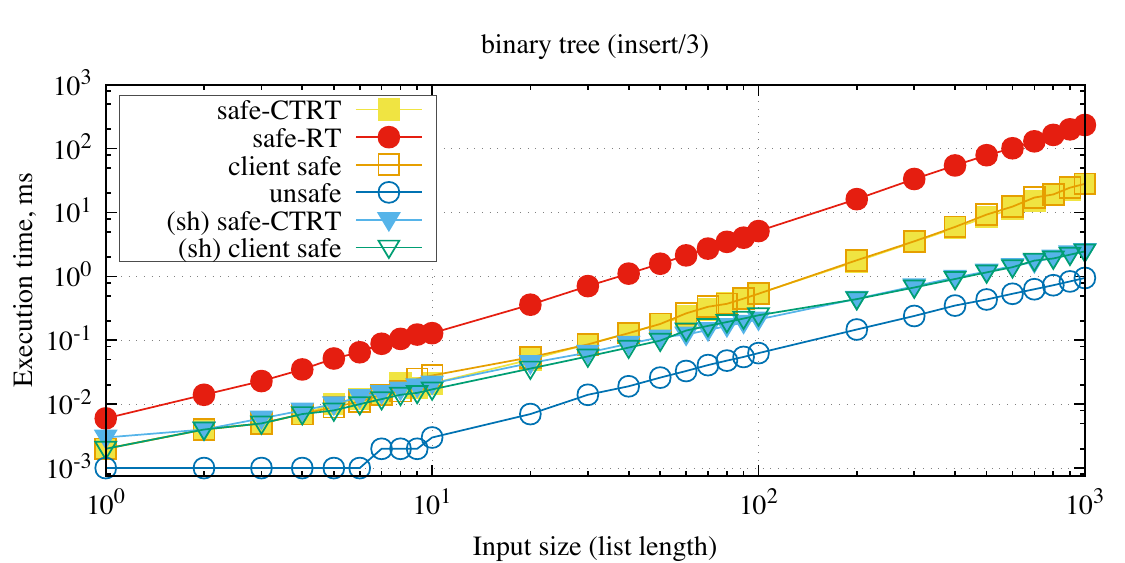}
\caption{Run times for the \kbd{binary tree} benchmark in different %
             execution modes, $O(log(N))$ operation + $O(N)$ check complexity}
    \includegraphics[width=\textwidth]{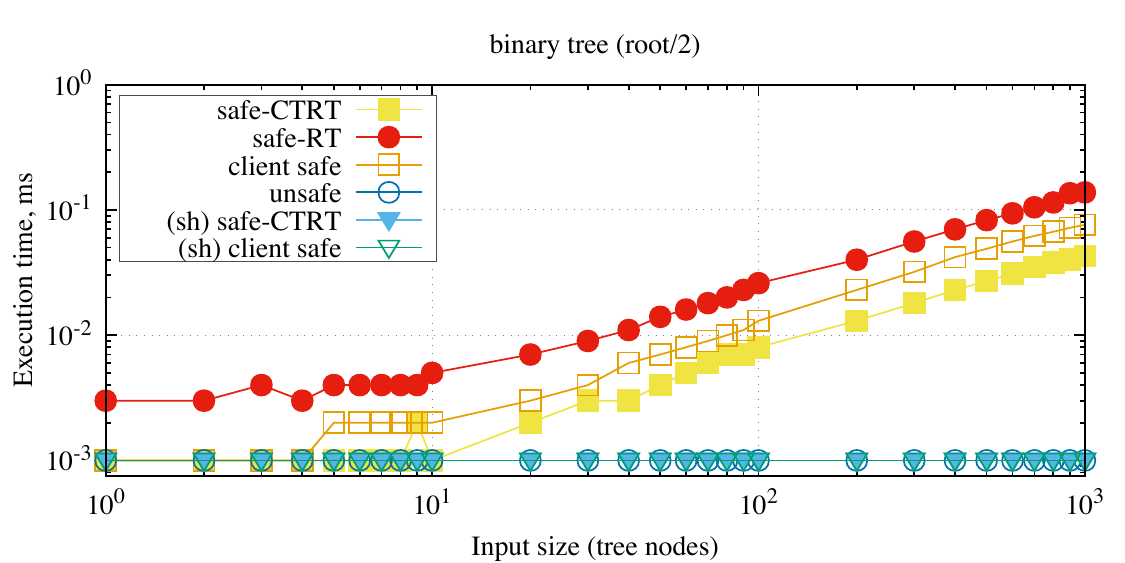}
\caption{Run times for the \kbd{binary tree} benchmark in different %
             execution modes, $O(1)$ operation + $O(N)$ check complexity}
\end{figure}
\vspace*{\fill}

\begin{figure}[t]
    \centering
    \includegraphics[width=\textwidth]{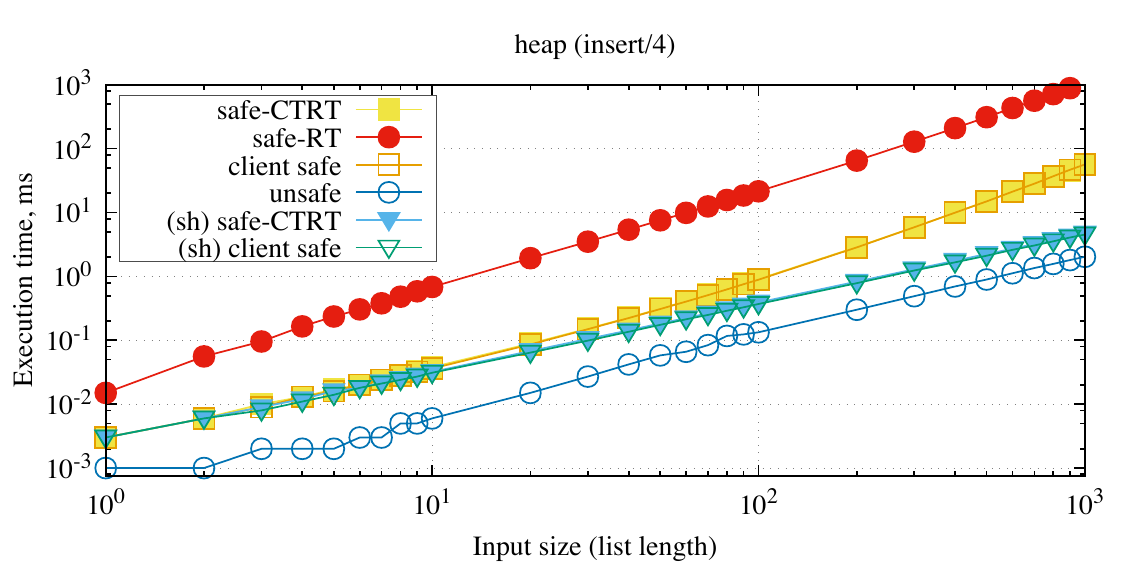}
\caption{Run times for the \kbd{min-heap} benchmark in different %
             execution modes, $O(log(N))$ operation + $O(N)$ check complexity}
    \includegraphics[width=\textwidth]{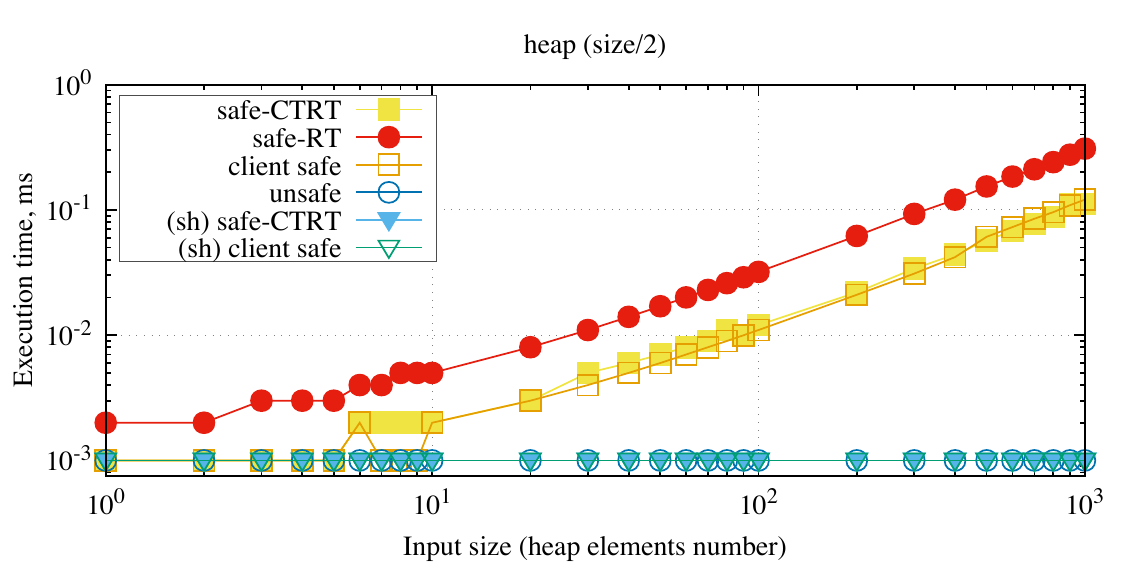}
\caption{Run times for the \kbd{min-heap} benchmark in different %
             execution modes, $O(1)$ operation + $O(N)$ check complexity}
\end{figure}

\begin{figure}[t]
    \centering
    \includegraphics[width=\textwidth]{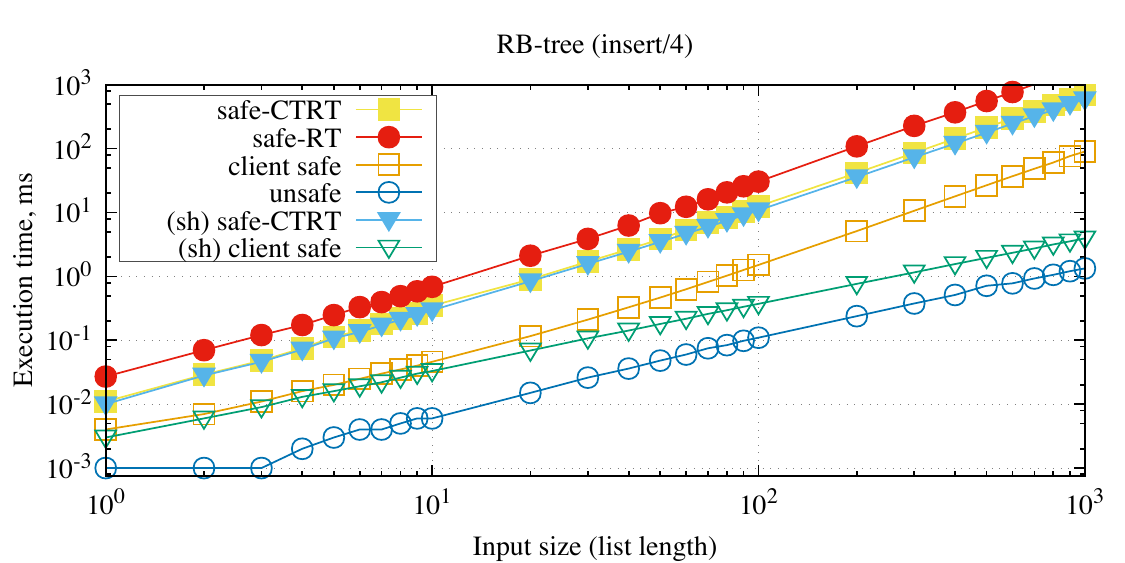}
\caption{Run times for the \kbd{RB-tree} benchmark in different %
             execution modes, $O(log(N))$ operation + $O(N)$ check complexity}
    \includegraphics[width=\textwidth]{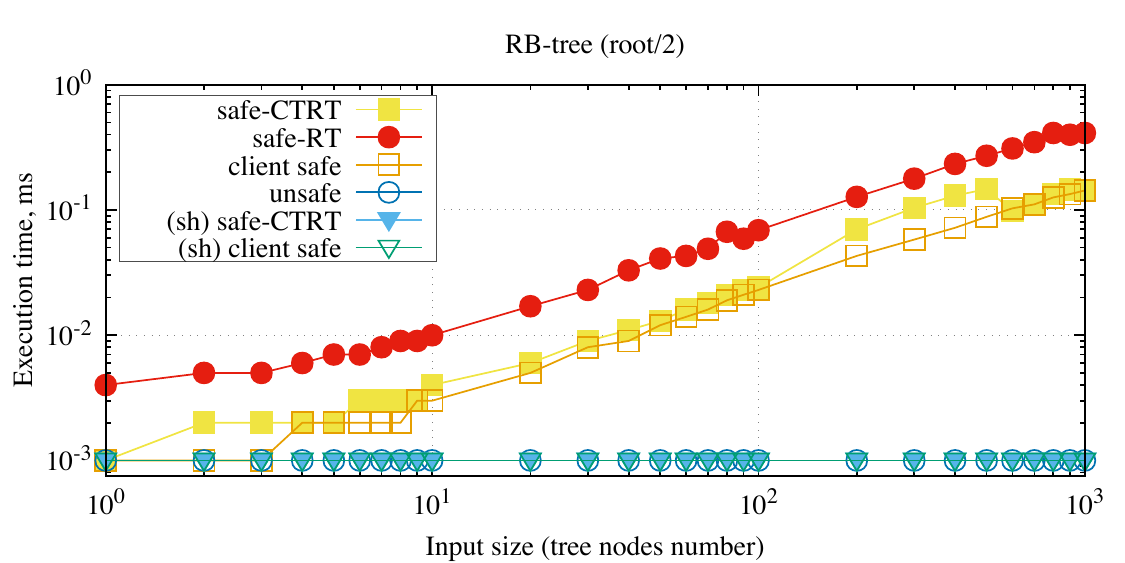}
\caption{Run times for the \kbd{RB-tree} benchmark in different %
             execution modes, $O(1)$ operation + $O(N)$ check complexity}
\end{figure}

\end{document}